\documentclass[a4paper,12pt]{article} 

\usepackage[a4paper,left=2.5cm,right=2.5cm,top=2.5cm,bottom=2.5cm]{geometry}

\usepackage[utf8]{inputenc}
\usepackage[T1]{fontenc}
\usepackage[english]{babel}

\usepackage{amsmath,amssymb}

\usepackage{graphicx}
\graphicspath{{Figures/}}
\usepackage{tabularx}
\usepackage{makecell}
\usepackage{array}
\usepackage{booktabs}
\usepackage{multirow}
\usepackage{float}
\usepackage{subcaption}
\usepackage{adjustbox}

\usepackage{enumitem}
\newlist{questions}{itemize}{1}
\setlist[questions]{label=\textbf{Q:}}

\usepackage{algorithm}
\usepackage[noend]{algpseudocode}
\usepackage{caption}
\usepackage{listings}
\usepackage{xcolor}
\definecolor{lightgray}{gray}{0.95}
\lstdefinestyle{pythoncode}{
    language=Python,
    backgroundcolor=\color{lightgray},
    basicstyle=\ttfamily\scriptsize,
    keywordstyle=\color{blue},
    commentstyle=\color{gray},
    stringstyle=\color{purple},
    showstringspaces=false,
    breaklines=true,
    tabsize=2
}

\usepackage[
    backend=biber,
    uniquename=false,
    style=authoryear,
    sorting=nyt,
    maxcitenames=2
]{biblatex}
\usepackage{csquotes}
\usepackage{authblk}
\usepackage[hidelinks]{hyperref}
\newif\ifdirty
\dirtyfalse % set \dirtytrue for highlighted changes
\ifdirty
  \newcommand{\chg}[1]{\hl{#1}}
\else
  \newcommand{\chg}[1]{#1}
\fi

\newcommand\blfootnote[1]{%
  \begingroup
  \renewcommand\thefootnote{}\footnote{#1}%
  \addtocounter{footnote}{-1}%
  \endgroup
}

\title{Stochastic Volatility Modelling with LSTM Networks: \\ 
A Hybrid Approach for S\&P 500 Index Volatility Forecasting}

\date{\vspace{-5ex}} % suppress date

\author[1]{Anna Perekhodko}
\author[2]{Robert Ślepaczuk}

\affil[1]{\small University of Warsaw, Faculty of Economic Sciences,\\
Ul. Długa 44/50, 00-241 Warsaw, Poland, email: anna.perekhodkoa@gmail.com}

\affil[2]{\small University of Warsaw, Faculty of Economic Sciences,\\
Department of Quantitative Finance and Machine Learning,\\
Quantitative Finance Research Group, Ul. Długa 44/50, 00-241 Warsaw, Poland,\\
ORCID: \url{https://orcid.org/0000-0001-5227-2014}, Corresponding author: rslepaczuk@wne.uw.edu.pl }

\begin{document}

\maketitle

\blfootnote{All code and supplementary materials used in this paper are 
openly accessible via GitHub: 

\url{https://github.com/aperekhodko/sv_lstm_hybrid_model}.}

\begin{abstract}
Accurate volatility forecasting is essential in various domains, including banking, investment, and risk management, as expectations about future market movements directly influence current decision-making. This study proposes a hybrid modeling framework that integrates a Stochastic Volatility model with a Long Short-Term Memory neural network. The SV model contributes statistical precision and the ability to capture latent volatility dynamics, particularly in response to unforeseen events, while the LSTM network enhances the model’s ability to detect complex, nonlinear patterns in financial time series. The forecasting is conducted using daily data from the S\&P 500 index, covering the period from January 1, 1998, to December 31, 2024. A rolling window approach is employed to train the model and generate one-step-ahead volatility forecasts. The performance of the hybrid SV-LSTM model is evaluated through both statistical testing and investment simulations. Results show that the hybrid approach outperforms both the standalone SV and LSTM models. These findings contribute to the ongoing development of volatility modeling techniques and provide a robust foundation for enhancing risk assessment and strategic investment planning in the context of the S\&P 500.
\end{abstract}

\section{Introduction}

The demand for accurate volatility forecasts reflects their central role in risk management, asset pricing, and portfolio optimization. Financial market participants rely on these forecasts to guide investment strategies and assess risk exposure. While models such as GARCH and SV capture the stochastic nature of volatility, incorporating LSTM networks can enhance predictions by identifying non-linear patterns. This study addresses a gap in financial forecasting research by proposing a hybrid SV-LSTM model that leverages both the statistical strengths of SV and the pattern-recognition capacity of LSTM.

The objective is to assess whether incorporating latent volatility estimates from an SV model as an additional input enhances the predictive performance of the LSTM model compared to their standalone applications.

The main focus of the forecast was chosen to be the volatility of the S\&P 500 index, a well-established measure of market fluctuations. 

The research conducted in this paper is guided by the following central hypotheses:
\begin{itemize}
  \item H1: The inclusion of stochastic volatility forecasts for day t+1 enhances the predictive accuracy of the LSTM model.
  \item H2: Augmenting the input data of the LSTM model with external information beyond historical returns improves its forecasting performance.
  \item H3: The hybrid SV-LSTM model delivers enhanced volatility forecasts compared to the standalone SV model.
\end{itemize}
In addition to the main hypotheses, the following secondary research questions are investigated:
\begin{itemize}
  \item RQ1:  Increasing the dimensionality of inputs from the SV model further improves the predictive performance of the hybrid model.
  \item RQ2: The preprocessing and transformation of input data have a significant effect on the performance of the SV-LSTM model.
  \item RQ3: The decreased sequence of the input data into the LSTM model improves the SV-LSTM prediction accuracy.

\end{itemize}

The data used is daily close price for the S\&P 500 index, covering the period from January 1, 1998, to December 31, 2024, obtained from the YahooFinance.

The methodology employed in this thesis involves a combination of stochastic volatility modeling, LSTM networks and statistical testing. The SV model is used to generate volatility forecasts, which are then incorporated as inputs to the LSTM model. Additionally, benchmarks, including standalone LSTM and SV models, are introduced to compare the performance of the hybrid SV-LSTM model. The Wilcoxon Signed-Rank and Diebold-Mariano Tests are used to statistically evaluate the significance of the differences in predictive performance between the models.

This study contributes to financial modeling by examining the combination of a statistical SV model as input to a machine learning LSTM model. It is the first to investigate the impact of incorporating SV predictions of latent volatility into the LSTM architecture. This hybrid approach captures both latent stochastic processes and nonlinear dependencies in financial time series, improving the model’s ability to reflect unpredictable market dynamics. The step-by-step framework, supported by sensitivity analysis and statistical testing, bridges the gap between deep learning techniques and traditional econometric approaches. Practically, the study demonstrates relevance through a simulated investment strategy, highlighting potential improvements in portfolio risk management, dynamic asset allocation, and algorithmic trading, and offering tools for more adaptive and resilient investment strategies in volatile markets.

The structure of the thesis is as follows: The second chapter provides a review of literature on both classical volatility models and hybrid financial forecasting. The third chapter presents the dataset used, preprocessing techniques and error metrics. The fourth chapter outlines the methodology for three models involved in the model development: SV model, LSTM and hybrid SV-LSTM. The fifth chapter compares the models’ performance using statistical tests, while the sixth chapter explores the robustness of the results through sensitivity analysis. Finally, the seventh chapter summarizes the key findings, discusses their implications, and reflects on the achievement of the research objectives.

\section{Literature Review}
The early development of financial econometric models was founded on the assumption of constant volatility,  which was most notably formalized in the seminal work of \textcite{blackscholes1973}. Their option pricing model introduced a cornerstone framework that assumed volatility to be constant over time, providing analytical tractability and laying the foundation for modern financial derivatives pricing. However, when confronted with empirical evidence from financial markets and clustering behavior, this simplifying assumption soon revealed substantial limitation.

The assumption of constant volatility was first formally challenged by \textcite{engle1982}, who introduced the Autoregressive Conditional Heteroskedasticity (ARCH) model. This framework allowed volatility to evolve as a function of past squared returns, capturing the fact of volatility clustering observed in asset returns. Based on this idea, \textcite{bollerslev1986} extended the model into the Generalized Autoregressive Conditional Heteroskedasticity (GARCH) specification, which incorporated both lagged squared returns and past conditional variances, thereby improving the model's flexibility and empirical performance.

Despite the empirical success of ARCH-type models, \textcite{engle2001} later critically examined their explanatory power, acknowledging that although these models effectively capture volatility clustering, they are insufficient in explaining the underlying sources of volatility dynamics. \textcite{nelson2001} further argued that ARCH models impose parameter restrictions that are often violated by estimated coefficients, potentially unduly restricting the dynamics of the conditional variance process. These critiques motivated the development of alternative approaches where volatility is modeled as an unobserved, latent process.

One such approach is the Stochastic Volatility (SV) model introduced by \textcite{taylor1982,taylor1986}, which treats volatility as a latent stochastic process rather than a deterministic function of past returns, as in GARCH. This model offers greater flexibility in capturing nonlinear behavior, though estimation remains challenging. Building upon the stochastic volatility framework, \textcite{heston1993} introduced a continuous-time model that incorporated stochastic variance into the option pricing formula. The Heston model extended the tractability of Black-Scholes by deriving a closed-form solution for option prices while allowing for mean-reverting stochastic volatility and, particularly,  allowing correlation between volatility and asset returns—an essential feature for capturing the so-called leverage effect. Further advancements were made by \textcite{kim1998},  addressing computational challenges in SV models employing Markov Chain Monte Carlo (MCMC) methods. These methods enabled efficient Bayesian estimation without direct likelihood evaluation by introducing strong tools for filtering, diagnostic checking, and formal model comparison. The superiority of SV models over GARCH-type models was further demonstrated by \textcite{yu2002} in an empirical study on the volatility of New Zealand stock market data.

The SV model was criticized for failing to reflect long-term dependencies in volatility dynamics despite its benefits. To address this limitation, \textcite{breidt1998} introduced the Long Memory Stochastic Volatility model, incorporating an Autoregressive Fractionally Integrated Moving Average (ARFIMA) process into the standard stochastic volatility framework. \textcite{kilic2011} further contributed to this discussion by proposing the Smooth Transition FIGARCH model, which accounts for both long memory and nonlinear dynamics in conditional variance.

Recurrent neural networks (RNNs) have become very effective tools for volatility forecasting in recent years. One particularly effective model is the Long Short-Term Memory (LSTM) network, introduced by \textcite{hochreiter1997}. LSTM networks consist of multiple layers and adjust their weights based on observed loss between predicted and actual values. The application of LSTMs in financial volatility forecasting gained traction with \textcite{bahadori2019}, who demonstrated that LSTMs provided accurate forecasts for the S\&P 500 index, supported by backtesting evidence. This line of research was extended by \textcite{bucci2020}, who validated the usability of neural networks for forecasting the logarithm of realized volatility while incorporating macroeconomic determinants. LSTM and Nonlinear Autoregressive with Exogenous Inputs (NARX) neural networks emerged as the best-performing models in this study.

Further empirical validation was provided by \textcite{michankow2022} in forecasting BTC and S\&P 500 index volatility. \textcite{grudniewicz2023} expanded this research by comparing multiple machine learning approaches in algorithmic investment strategies. Their study evaluated models including Neural Networks, K-Nearest Neighbors, Regression Trees, Random Forests, Naïve Bayes Classifiers, Bayesian Generalized Linear Models, Support Vector Machines, and Linear Support Vector Machines, with Bayesian Generalized Linear Models and Linear SVMs exhibiting the best predictive performance.

The incorporation of LSTM models into hybrid volatility modeling techniques has been powered by their success.   \textcite{kim2018} proposed an LSTM-GARCH hybrid model applied to the volatility of the KOSPI 200 stock index, with the EGARCH-LSTM variant yielding the best performance. \textcite{rozynska2024} further explored this hybrid approach by incorporating the VIX index as an additional input in the GARCH-LSTM model. Meanwhile, \textcite{nguyen2019} extended LSTM applications to the SV framework, addressing the short-term memory limitations of SV models and demonstrating strong performance in forecasting weekly index data for S\&P 500 and ASX 200 using the Blocking Pseudo Method and Importance Sampling Squared algorithm.

Pursuing this trajectory, the current thesis suggests a distinctive enhancement to the SV framework by combining it with LSTM. Specifically, the aim is to simulate the joint posterior distribution of the SV parameters using MCMC methods. This approach will generate MCMC draws to produce one-step-ahead predictions ($t+1$) of returns and volatilities. These outputs will then be integrated into an LSTM architecture, leveraging its capacity to model long-term dependencies and nonlinear dynamics. By combining SV’s stochastic flexibility with LSTM’s predictive power, this hybrid model seeks to address the limitations of standalone SV models, such as their short-term memory bias and improve forecasting accuracy for financial time series. By providing a comprehensive framework for volatility modeling and prediction, this contribution attempts to narrow the gap between the empirical adaptability of machine learning and the theoretical robustness of stochastic volatility.

\section{Data and Methodology}
\subsection{Data Inputs}
For the modeling of S\&P 500 index volatility, the dataset consisted of daily closing prices obtained from Yahoo Finance, spanning January 1, 1998, to December 31, 2024. This period spanned across multiple economic cycles, including the dot-com bubble, the 2008 financial crisis, and the post-COVID market recovery, providing a rich basis for volatility analysis. From these closing prices, daily log returns were calculated to normalize price movements and facilitate comparisons across time. The log return at time \( t \) was computed as:

\begin{equation}
r_t = \ln \left( \frac{P_t}{P_{t-1}} \right),
\end{equation}

where \( P_t \) denoted the closing price of the S\&P 500 index on day \( t \) and  \( P_{t-1} \) on day \(t-1\). This transformation was preferred in financial analyses because it converted absolute price changes into relative terms, mitigating the impact of scale and enabling stationarity in the return series, a critical assumption for volatility modeling.

Table~\ref{tab:descriptive_stats} summarized the descriptive statistics of the close prices and log returns over the 27-year period, offering insights into their distributional properties. The close prices exhibited a mean of 2044.98, significantly higher than the median of 1432.73, which suggested a right-skewed distribution. This discrepancy indicated that extremely high values pulled the mean upward. The standard deviation of 1220.20 reinforced this observation, reflecting substantial variability in daily prices and pointing to the volatile nature of the S\&P 500 over the sample period. The minimum price of 676.53, observed during the 2008--2009 financial crisis, and the maximum of 6090.27, likely from late 2024, further highlighted the dataset's wide range. A skewness of 1.3333 confirmed the positive skew, while a kurtosis of 0.8033 indicated a distribution flatter than a normal curve, suggesting fewer extreme deviations than expected under normality.

In contrast, the log returns displayed a mean of 0.000265 and a median of 0.000633, with the closeness of these central measures suggesting a more symmetric distribution than the close prices. However, the standard deviation of 0.012225 underscored significant day-to-day fluctuations, consistent with the high volatility typical of equity indices. The minimum log return of -0.127652 and the maximum of 0.109572, illustrated the potential for extreme movements. A negative skewness of -0.3822 indicated a slight leftward tilt, implying more frequent small negative returns than large positive ones---a common feature in financial time series. The kurtosis of 9.8617 was notably high, signaling a leptokurtic distribution with fat tails. This excess kurtosis suggested a higher-than-normal probability of extreme events, such as market crashes or rapid rallies, which was critical for volatility modeling as it underscored the need for models capable of capturing these tail risks.

\begin{table}[h!]
    \centering
    \renewcommand{\arraystretch}{1.4} % Adjust row spacing
    \setlength{\tabcolsep}{8pt}      % Adjust column spacing
    \normalsize
    \caption{Descriptive Statistics for S\&P 500 Close Prices and Log Returns (January 1, 1998 -- January 1, 2025)}
    \begin{tabularx}{\textwidth}{XXX}
        \hline
        \textbf{Statistic} & \textbf{Close Prices} & \textbf{Log Returns} \\
        \hline
        Mean       & 2044.98  & 0.000265  \\
        Median     & 1432.73  & 0.000633  \\
        Std Dev    & 1220.20  & 0.012225  \\
        Min        & 676.53   & -0.127652 \\
        Max        & 6090.27  & 0.109572  \\
        Skewness   & 1.3333   & -0.3822   \\
        Kurtosis   & 0.8033   & 9.8617    \\
        \hline
    \end{tabularx}
 \label{tab:descriptive_stats}
 
    \vspace{0.5em}
{\scriptsize
\noindent\parbox{\textwidth}{
Note: The following table presents the descriptive statistics of the S\&P 500 Index's closing prices and their corresponding log returns. A key observation is the relatively smaller distance between the mean and median in the distribution of log returns compared to that of the closing prices. This suggests that the distribution of log returns is more symmetric and potentially closer to a normal distribution. In contrast, the closing prices exhibit greater skewness
}
}
\end{table}

\subsection{Estimation of Volatility}
In this study, the forecasting target was established as the rolling historical volatility, which was computed over a window of \( N = 21 \) days, approximating one trading month. This choice of window length was selected to strike a balance between capturing short-term market dynamics and providing a stable estimate of volatility, a common practice in financial time series analysis. The rolling volatility was estimated using the unbiased standard deviation formula:

\begin{equation}
\sigma_t = \sqrt{\frac{1}{N-1} \sum_{i=t-N+1}^{t} (r_i - \bar{r}_t)^2},
\end{equation}

where \( \sigma_t \) is the rolling volatility at time \( t \), \( N \) is the window size over which volatility is calculated. The denominator \( N-1 \) in the volatility formula provides an unbiased estimate of the standard deviation. Meanwhile, \( r_i \) represents the logarithmic return at time \( i \) and \( \bar{r}_t \) is the rolling mean of the logarithmic returns within the window, given by:

\begin{equation}
\bar{r}_t = \frac{1}{N} \sum_{i=t-N+1}^{t} r_i.
\end{equation} 

Here, \( \bar{r}_t \) serves as the average return over the window, ensuring that deviations \( (r_i - \bar{r}_t) \) measure fluctuations around the mean. 

\begin{figure}[H]
    \centering
    \caption{S\&P 500 21-Day Rolling Volatility Over Time (1998--2025)}
    \includegraphics[width=0.8\textwidth]{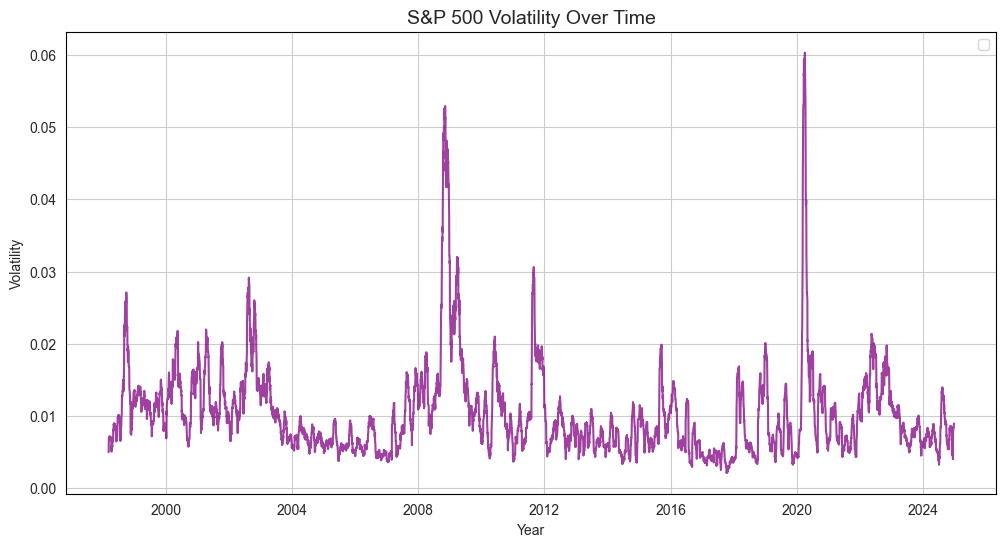}
    \label{fig:vol_time}

\vspace{0.5em}
{\scriptsize
\noindent\parbox{\textwidth}{
 Note: The figure provides the time-series of rolling volatility over time for the time frame from January 1998 to December 2024.
}
}
\end{figure}

This approach ensured that the volatility measure reflected recent market conditions while accounting for the sample variance’s degrees of freedom, enhancing its statistical robustness. The 21-day window was particularly suitable for the S\&P 500, as it aligned with monthly trading cycles and captured key volatility clustering patterns observed in the log returns, such as those during the 2008 financial crisis and the 2020 COVID-19 market turmoil, which can be observed in more detail in Figure ~\ref{fig:vol_time}. This rolling historical volatility served as the benchmark for evaluating the predictive accuracy of the Stochastic Volatility (SV), LSTM, and Hybrid SV-LSTM models, as detailed in subsequent sections.

\subsection{Data Preprocessing}

For the LSTM model, data preprocessing involved scaling the log returns to ensure numerical stability and compatibility with the neural network’s optimization process. The Min-Max normalization formula was applied as follows:

\begin{equation}
X_t^{\text{scaled}} = \frac{X_t - \min(X)}{\max(X) - \min(X)} \cdot (1 - 10^{-11}) + 10^{-11},
\end{equation}

Where \( X \) denotes the set of log return values within each rolling window segment. The small constant \( 10^{-11} \) was incorporated to prevent division-by-zero issues and ensure that the scaled values ranged between \( 10^{-11} \) and \( 1 \). The \chg{max(}\(X\)\chg{) and min(}\(X\)\chg{) reflected the maximum and minimum values from each respective dataset.}

To safeguard against data leakage scaling was performed in two distinct steps. First, the training and validation sets were scaled together to establish a consistent range for model tuning. Then, the training set alone was scaled separately, ensuring that the maximum and minimum values used for scaling did not incorporate information from the test set. After predictions were generated, the scalers applied to the test data were stored and utilized to inverse-transform the predicted values back to their original scale, enabling a direct and meaningful comparison with the actual rolling volatility.

In contrast, the SV model required no scaling, as it operated directly on the raw log returns, leveraging its statistical framework to model volatility through latent processes. 

This preprocessing strategy ensured that each model received appropriately prepared inputs tailored to its methodological requirements, maintaining consistency across the analysis.

\subsection{Evaluation Metrics}

The choice of error metrics was crucial, serving both as objective functions for model training and as benchmarks for evaluating performance across the SV, LSTM, and Hybrid SV-LSTM models. Three widely used metrics: Mean Squared Error (MSE), Mean Absolute Error (MAE) and Mean Absolute Percentage Error (MAPE) were selected for their complementary strengths in capturing different dimensions of forecast accuracy, well suited to volatility prediction.

\subsubsection*{Mean Squared Error (MSE)}

The Mean Squared Error (MSE) quantified the average squared difference between the actual rolling volatility values and the predicted values. It was calculated as:

\begin{equation}
\text{MSE} = \frac{1}{n} \sum_{i=1}^{n} (y_i - \hat{y}_i)^2,
\end{equation}

where \( y_i \) represented the actual volatility, \( \hat{y}_i \) denoted the predicted volatility, and \( n \) was the number of observations in each rolling window. %/The squaring of differences increased the significance of larger errors, penalizing significant deviations.

\subsubsection*{Mean Absolute Error (MAE)}

The Mean Absolute Error (MAE) measured the average magnitude of errors between actual and predicted volatility, irrespective of their direction. It was defined as:

\begin{equation}
\text{MAE} = \frac{1}{n} \sum_{i=1}^{n} |y_i - \hat{y}_i|,
\end{equation}

%where the variables retained the same meanings as in MSE. Unlike MSE, MAE did not disproportionately penalize larger errors, providing a more balanced assessment of performance across all predictions. 

\subsubsection*{Mean Absolute Percentage Error (MAPE)}

The Mean Absolute Percentage Error (MAPE) assessed the error relative to the actual volatility, expressing it as a percentage. It was computed as:

\begin{equation}
\text{MAPE} = \frac{1}{n} \sum_{i=1}^{n} \left| \frac{y_i - \hat{y}_i}{y_i} \right| \times 100,
\end{equation}

%With variables consistent with the previous metrics. MAPE normalized the absolute error by the magnitude of the actual values, offering insight into the proportional accuracy of predictions. 

The combination of MSE, MAE, and MAPE was deliberate, reflecting their ability to address distinct facets of volatility prediction.  

\subsection{Modeling Framework}

This study aimed to forecast the one-day-ahead (\( t + 1 \)) volatility of the S\&P 500 index. To achieve this goal, three models were explored: the Stochastic Volatility (SV) model, the Long Short-Term Memory (LSTM) neural network and a Hybrid SV-LSTM model that combined elements of both approaches.

Each model utilized the dataset described in Section 3.1 and implemented a rolling window methodology to capture the time-varying nature of financial volatility over the period from January 1, 1998, to January 1, 2025. Table~\ref{tab:model_summary} provides a summary of the inputs, time frames, and outputs for each model.

\begin{table}[H]
\centering
\caption{Summary of Models, Inputs, and Outputs}
\label{tab:model_summary}
\renewcommand{\arraystretch}{1.2}
\resizebox{\textwidth}{!}{%
\begin{tabular}{c >{\centering\arraybackslash}p{5cm} c >{\centering\arraybackslash}p{4.5cm}} \hline 
\textbf{Model} & \textbf{Inputs} & \textbf{Period} & \textbf{Output} \\ \hline 
SV & Log returns & 1998--2025 & Latent volatility at \( t+1 \) \\ 
LSTM & Log returns, 21-day historical volatility & 2000--2025 & Rolling volatility at \( t+1 \) \\ 
Hybrid SV-LSTM & SV volatility at \( t+1 \), log returns, 21-day historical volatility & 2000--2025 & Rolling volatility at \( t+1 \) \\ \hline
\end{tabular}
}
\vspace{0.5em}
{\scriptsize
\noindent\parbox{\textwidth}{Note: The table provides a summary of each model used in the study, detailing their respective input variables, the period over which the input data is drawn, and the specific output of the model.}
}
\end{table}

The SV model represents a statistical approach in which volatility is modeled as a latent stochastic process driven by the log returns of the S\&P 500 index. This model was selected for its flexibility in capturing time-varying volatility and heavy-tailed distributions typical in financial data. The complete historical dataset from 1998 to 2025 was used, with a rolling window of 504 trading days (approximately 2 years). For each step, the window was moved forward by one trading day to produce a new one-step-ahead forecast, thus adapting to changes in the data over time.

In contrast, the LSTM model adopted a deep learning approach, known for its capacity to learn long-range and nonlinear patterns in sequential data. The model used data from 2000 to 2025 and was fed \chg{with} the log returns and the 21-day rolling historical volatility as inputs. Its objective was to forecast the same rolling volatility for day (\( t+1 \)).

\begin{figure}[H]
\centering
\caption{Schematic of the rolling window methodology used for LSTM and Hybrid models}
\includegraphics[width=0.8\textwidth]{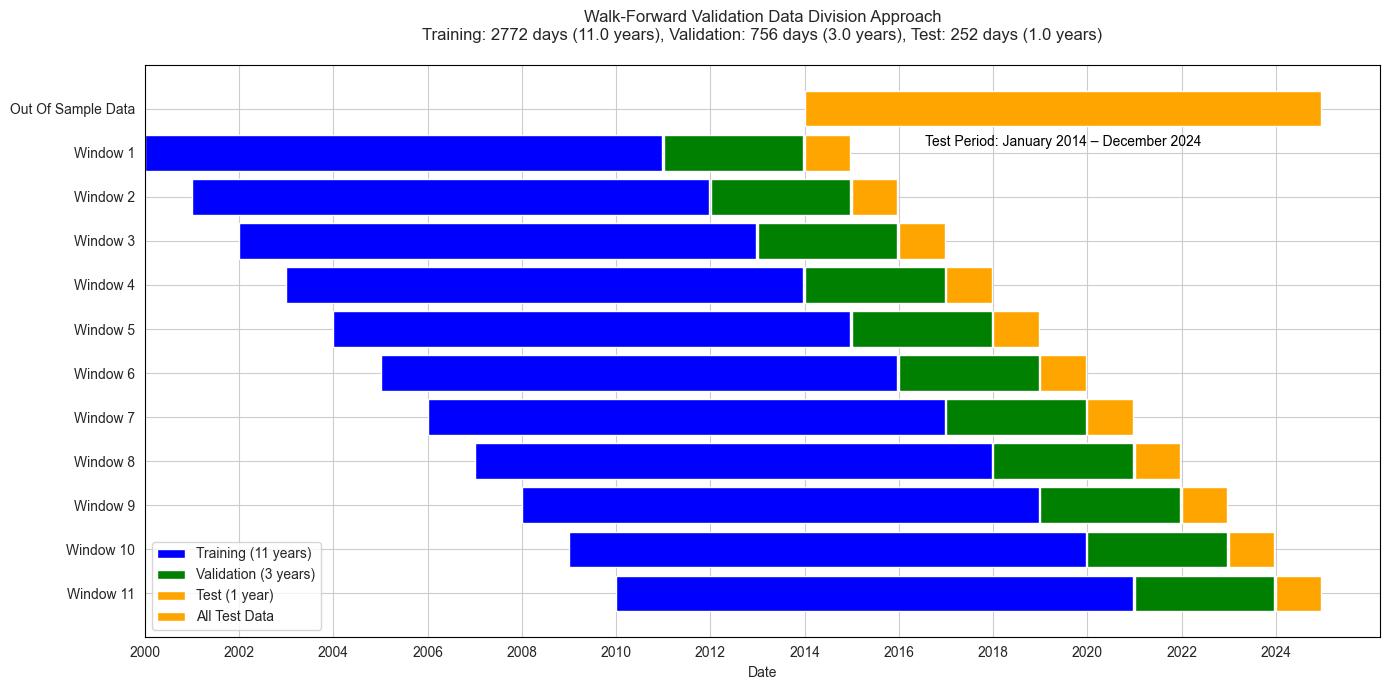}
\label{fig:rolling_window}

\vspace{0.5em}
{\scriptsize
\noindent\parbox{\textwidth}{
Note: The scheme illustrates the division of data into \chg{training, validating, and testing} sets for the LSTM model. Each window contains 11 years of training data, 3 years of validation data, and 1 year of test data. The window moves forward by one year at a time, and the model is refitted at each step. As a result, the out-of-sample data comprises the test segments from 11 consecutive windows, collectively forming the final test set. Predictions were generated for the next day (\( t + 1 \)) in each case.
}
}
\end{figure}

To accommodate the training requirements of neural networks, a multi-year rolling window approach was used, as illustrated on Figure~\ref{fig:rolling_window}):

\begin{itemize}
    \item Each window comprised 11 years of training data (approximately 2772 trading days), 3 years of validation data (756 trading days), and 1 year of test data (252 trading days).
    \item After each test period, the window shifted forward by one year (252 trading days), creating 11 overlapping windows across the 25-year dataset.
\end{itemize}

This setup ensured the model was trained on extensive historical data, validated for generalization, and tested on out-of-sample periods — closely resembling real-world forecasting settings.

The Hybrid SV-LSTM model was constructed to integrate the statistical accuracy of the SV model with the learning capacity of the LSTM. It combined three inputs: the SV model’s latent volatility prediction for \( t+1 \), the log returns, and the 21-day rolling historical volatility. Like the LSTM model, it used data from 2000 to 2025 and followed the same rolling window structure. By incorporating SV predictions as an additional input, the model was designed to enrich the LSTM's feature space with statistically derived volatility signals.

\section{Model Development}
\subsection{Stochastic Volatility Model}
%\The stochastic volatility (SV) model provides a robust and theoretically grounded framework for modeling financial time series, particularly when volatility exhibits time-varying and unpredictable behavior, as is common in equity indices like the S\&P 500. 

First introduced by Taylor [1982, 1986], the SV model diverges from traditional deterministic volatility models, such as the Generalized Autoregressive Conditional Heteroskedasticity (GARCH) framework [Bollerslev, 1986], by treating volatility as a latent stochastic process rather than a deterministic function of past observations. This approach offers greater flexibility in capturing empirical features of financial returns, such as volatility clustering, mean reversion, and the leverage effect, making it preferable to GARCH for this study, especially given evidence of its superior performance in modeling complex volatility dynamics. 
\subsubsection{Model Selection}

In the Stochastic Volatility (SV) framework, asset returns are modeled as conditionally normally distributed with a time-varying variance driven by a latent log-volatility process, as implemented in the \texttt{stochvol} package (Kastner, 2016). The mathematical formulation is expressed through the following state-space equations:

\begin{itemize} 
    \item \textbf{Return Equation:}
    \begin{equation}
    y_t \mid h_t \sim N(0, e^{h_t})
    \end{equation}
    where \( y_t \) is the observed return at time \( t \), \( h_t \) is the latent log-volatility, and \( e^{h_t} \) gives the variance at time \( t \).

    \item \textbf{Latent Volatility Process:}
    \begin{equation}
    h_t \mid h_{t-1}, \mu, \phi, \sigma_\eta \sim N\!\left(\mu + \phi (h_{t-1} - \mu), \sigma_\eta^2\right)
    \end{equation}
    where \( h_t \) is the log-volatility at time \( t \), \( h_{t-1} \) is its value at the previous time step, \( \mu \) is the long-term mean of the log-volatility, \( \phi \) is the persistence parameter, and \( \sigma_\eta \) is the volatility of the volatility process.  This equation models the evolution of log-volatility over time, where the process exhibits mean-reverting behavior and can experience random shocks due to volatility clustering.

    \item \textbf{Initial Condition:}
    \begin{equation}
    h_0 \mid \mu, \phi, \sigma_\eta \sim N\!\left(\mu, \frac{\sigma_\eta^2}{1 - \phi^2}\right)
    \end{equation}
    where \( h_0 \) is the initial log-volatility, \( \mu \) is the long-term mean, \( \phi \) is the persistence parameter, and \( \sigma_\eta \) is the volatility of the volatility process. This equation specifies the distribution of the initial value of log-volatility, ensuring that it is consistent with the subsequent evolution of volatility dynamics.
\end{itemize}

The stochastic specification allows volatility to evolve randomly over time while exhibiting mean-reverting behavior, a well-documented characteristic of financial markets (Engle \& Patton, 2001). This feature distinguishes the SV model from GARCH models and enhances its suitability for capturing the intricate volatility patterns observed in real-world data.

\subsubsection{Prior Parametrization and Sampling}

The SV model hinges on three key parameters: \( \mu \), the long-term mean of log-volatility; \( \phi \), the persistence of the volatility process; and \( \sigma_{\eta} \), the volatility of the volatility process, each of which requires careful parametrization and estimation. In this study, we adopt a data-driven approach by employing weakly informative default priors, as inspired by Kim et al. (1998), rather than imposing strong prior distributions, ensuring flexibility across forecasting windows and leveraging priors well-calibrated for general financial time series. Specifically, the priors are set independently for each parameter within the \texttt{stochvol} package (Kastner, 2016), relying on posterior distributions for inference to reflect the underlying data dynamics. Parameter estimation is conducted using Markov Chain Monte Carlo (MCMC) sampling via the Metropolis-Hastings algorithm, implemented efficiently in the \texttt{stochvol} package. This method iteratively samples from the posterior distributions of the latent log-volatility \( h_t \) and the parameters \( \mu \), \( \phi \), and \( \sigma_{\eta} \), utilizing the package’s ancillarity-sufficiency interweaving strategy (ASIS) to enhance mixing and convergence (Kastner \& Frühwirth-Schnatter, 2014). The sampling process runs for 1,000 iterations, with a 200-iteration burn-in period discarded to ensure the chain stabilizes, providing reliable estimates for volatility forecasting.

\subsubsection{SV Model Results}

The stochastic volatility (SV) model has been implemented to generate one-day-ahead median volatility forecasts for the S\&P 500, spanning February 1, 1997, to January 1, 2025, using a rolling window approach with a training period of 504 days, equivalent to two years of trading data. At this interim stage, the model produces predictions representing the median of the predictive distribution for \( \sqrt{e^{h_{t+1}}} \), where \( h_{t+1} \) denotes the latent log-volatility at time (\( t+1 \)), derived from the posterior draws of the volatility process as outlined in Subsection 4.1.1.

\begin{figure}[h!]
    \centering
    \caption{SV Model Forecasts vs. \chg{21-Day Rolling Historical Volatility}}
    \includegraphics[width=1\linewidth]{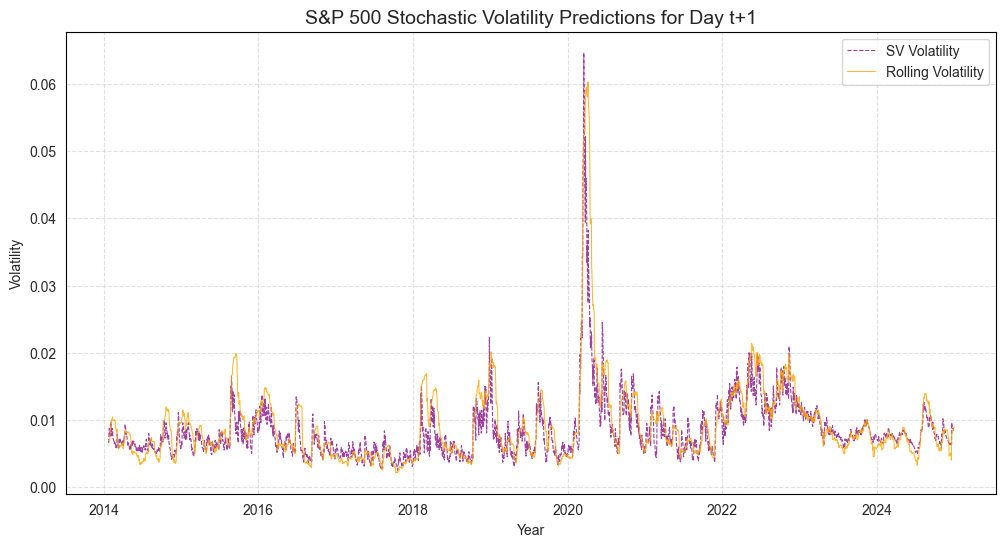}
    \label{fig:vol_comparison}

    \vspace{0.5em}
{\scriptsize
\noindent\parbox{\textwidth}{
Note: The figure shows the predicted latent volatility for day \(t+1\), generated by the Stochastic Volatility model, plotted against the \chg{21-day historical rolling volatility} for the period from January 2014 to December 2024.
}
}
\end{figure}

Focusing on January 24, 2014, to December 30, 2024, Figure~\ref{fig:vol_comparison} compares these forecasts to the two-year rolling historical volatility, revealing that SV estimates capture key trends, such as the 2020 spike and early 2022 fluctuations, but exhibit more noise than the smoother benchmark.

Table~\ref{tab:error_metrics} summarizes the error metrics. The MAPE of 18.12\% reflects moderate accuracy, influenced by the model’s sensitivity to short-term noise, while the MSE of \( 9 \times 10^{-6} \) and MAE of \( 1.717 \times 10^{-3} \) indicate small absolute errors.

\begin{table}[h!]
\centering
\caption{Comparative Error Metrics: SV Model vs. \chg{21-Day Rolling Historical Volatility} on Period January 2014 - December 2024}
\label{tab:error_metrics}
\begin{tabular}{lc}
\hline
\textbf{Metric} & \textbf{Value} \\
\hline
Mean Absolute Percentage Error (MAPE) & 18.12\% \\
Mean Squared Error (MSE) & \( 9 \times 10^{-6}\)  \\
Mean Absolute Error (MAE) & \(1.717 \times 10^{-3}\) \\
\hline
\end{tabular}

\vspace{0.5em}
{\scriptsize
\noindent\parbox{\textwidth}{
Note: The table presents the accuracy metrics calculated for the out-of-sample predictions of the Stochastic Volatility model for day \(t+1\), covering the period from January 2014 to December 2024.
}
}
\end{table}

\subsection{LSTM Model}
\subsubsection{Model Selection and Parameterization}
The Long Short-Term Memory (LSTM) model, \chg{originally proposed by} \textcite{hochreiter1997}, was selected for its adeptness at time series forecasting, particularly in capturing long-term dependencies and adapting to volatile shocks prevalent in financial market volatility, such as that of the S\&P 500. Unlike traditional recurrent neural networks (RNNs), which struggle with vanishing gradients and retaining distant information, the LSTM’s architecture leverages memory cells and three key gates- forget, input, and output to regulate information flow effectively. 

The forget gate in an LSTM, defined as $f_t = \sigma(U_f \cdot x_t + V_f \cdot h_{t-1} + b_f)$, where $h_{t-1}$ is the previous hidden state, $x_t$ is the current input, and $b_f$ is the bias vector, uses a sigmoid activation function to determine how much of the previous cell state $C_{t-1}$ to retain. The sigmoid outputs values between 0 and 1, and the filtered cell state is computed as 
\[ C'_t = f_t \cdot C_{t-1}, \]
where $f_t$ acts as a retention factor (0 for full forgetting, 1 for full retention).

Meanwhile, the input gate, computed as $i_t = \sigma(U_i \cdot x_t + V_i \cdot h_{t-1} + b_i)$, decides how much new information to add to the cell state. It works alongside a candidate state, defined as 
\[ C^+_t = \tanh(U_c \cdot x_t + V_c \cdot h_{t-1} + b_c), \]
which proposes potential updates using a tanh activation (ranging from $-1$ to 1). The cell state is then updated as 
\[ C_t = C'_t + i_t \cdot C^+_t, \]
combining the filtered previous state (from the forget gate) with the gated new information (from the input gate and candidate state).

The output gate, given by $o_t = \sigma(U_o \cdot x_t + V_o \cdot h_{t-1} + b_o)$, shapes the hidden state with the formula 
\[ h_t = o_t \cdot \tanh(C_t). \]
The sigmoid $o_t$ (0 to 1) filters the tanh-transformed cell state (scaled between $-1$ and 1), ensuring relevant volatility patterns are preserved and passed to the next time step or output layer. This framework operates across training, validation, and test phases, where the model processes input sequences through these gates to predict one-day-ahead volatility. During training, a selected loss function is minimized by adjusting parameters (weights and biases) based on the error between predictions and actual values. This adjustment uses gradient descent with the update rule 
\[ w_t = w_{t-1} - \eta \cdot \frac{\partial L}{\partial w}, \]
where $\eta$ is the learning rate, $\frac{\partial L}{\partial w}$ is the gradient of the loss with respect to the parameter, and the term $\eta \cdot \frac{\partial L}{\partial w}$ represents the step size for updates. 

To mitigate overfitting, validation loss is monitored in real-time; an increase in validation error despite decreasing training loss signals over-specialization, prompting early stopping. Data is partitioned into overlapping rolling windows, with training (11 years), validation (3 years), and test (1 year) sets kept independent, as detailed in Subsection 4.1. Within each window, hyperparameter tuning is conducted using 25 random search combinations. Each combination is tested across three trials, with each trial running for 50 epochs. Early stopping is implemented to halt the training process if the validation loss does not improve over 5 consecutive iterations. The best model weights, corresponding to the lowest validation loss, are retained. Additionally, the average validation loss from the three trials is calculated to ensure robustness and reduce the likelihood of overfitting or anomalous results.The hyperparameters tuned, listed in Table~\ref{tab:lstm_hyperparams}, include the number of LSTM layers, dense layers, units per layer, learning rate , activation functions, recurrent dropout, dropout and loss function. Recurrent dropout, applied to recurrent connections, differs from standard dropout on feedforward layers by stabilizing temporal dependencies, enhancing the LSTM’s ability to generalize across volatile financial data. \chg{The selected hyperparameter grid as a result of tunning process can be observed in Figure}~\ref{tab:hps_selected}.

\begin{table}[h!]
\centering
\caption{Hyperparameters Tuned for the LSTM Model}
\label{tab:lstm_hyperparams}
\begin{tabular}{ll}
\hline
\textbf{Hyperparameter} & \textbf{Range} \\
\hline
Number of LSTM Layers & 1, 2 , 3 \\
Number of Dense Layers & 0, 1 ,2 , 3 \\
Units per Layer & 32, 64, 128 \\
Learning Rate & \( [10^{-4}, 5 \times 10^{-4}, 10^{-3}, 5 \times 10^{-3}, 10^{-2}] \) \\
Activation Function & tanh, relu, sigmoid \\
Recurrent Dropout & 0, 0.05, 0.1, 0.15, 0.2 \\
Dropout & 0, 0.05, 0.1, 0.15, 0.2 \\
Loss Function & MSE, MAE \\
\hline
\end{tabular}

\vspace{0.5em}
{\scriptsize
\noindent\parbox{\textwidth}{
Note: The table presents the range of hyperparameters selected for tuning the architecture of the LSTM model. A Random Search approach was employed, consisting of 25 trials, each with 50 epochs and 3 executions per trial. Early stopping was applied with a patience of 5. The tuning process was conducted across all rolling windows using the training and validation sets.
}
}
\end{table}

\begin{table}[h!]
\centering
\scriptsize
\caption{Selected LSTM Hyperparameter Grid for Rolling Windows 1–11}
\label{tab:hps_selected}
\begin{adjustbox}{max width=\textwidth}
\begin{tabular}{lcccccccccccc}
\toprule
\textbf{Hyperparameter/Window} & \textbf{1} & \textbf{2} & \textbf{3} & \textbf{4} & \textbf{5} & \textbf{6} & \textbf{7} & \textbf{8} & \textbf{9} & \textbf{10} & \textbf{11} \\
\midrule
Learning Rate & 0.005 & 0.01 & 0.01 & 0.01 & 0.0005 & 0.01 & 0.001 & 0.005 & 0.01 & 0.01 & 0.0005 \\
Loss Function & \multicolumn{11}{c}{MSE} \\
Number of LSTM Layers & 1 & 1 & 3 & 3 & 2 & 2 & 2 & 1 & 2 & 1 & 1 \\
LSTM Units (Layer 1) & 32 & 64 & 64 & 32 & 64 & 32 & 64 & 32 & 64 & 32 & 128 \\
Activation (Layer 1) & tanh & sigmoid & relu & tanh & sigmoid & sigmoid & tanh & relu & sigmoid & relu & sigmoid \\
Recurrent Dropout (L1) & 0 & 0 & 0.1 & 0.2 & 0.05 & 0.1 & 0 & 0 & 0.05 & 0.15 & 0.15 \\
LSTM Units (Layer 2) & 64 & 128 & 32 & 128 & 64 & 128 & 128 & 64 & 32 & 64 & 32 \\
Activation (Layer 2) & sigmoid & sigmoid & tanh & tanh & tanh & sigmoid & sigmoid & relu & sigmoid & sigmoid & tanh \\
Recurrent Dropout (L2) & 0.15 & 0.1 & 0 & 0.15 & 0 & 0.05 & 0.05 & 0 & 0.1 & 0.1 & 0.05 \\
LSTM Units (Layer 3) & 64 & 128 & 32 & 32 & 64 & 32 & 32 & 64 & 128 & 64 & 128 \\
Activation (Layer 3) & tanh & tanh & tanh & sigmoid & tanh & tanh & sigmoid & sigmoid & tanh & sigmoid & relu \\
Recurrent Dropout (L3) & 0 & 0.15 & 0 & 0.2 & 0.2 & 0.2 & 0 & 0 & 0.1 & 0.1 & 0.15 \\
Use Dropout Layer & \multicolumn{11}{c}{False} \\
Number of Dense Layers & 3 & 1 & 2 & 3 & 1 & 0 & 0 & 3 & 3 & 1 & 0 \\
Dense Units (Layer 1) & 128 & 32 & 32 & 32 & 128 &  &  & 64 & 64 & 32 &  \\
Activation (Dense 1) & tanh & relu & tanh & tanh & tanh &  &  & relu & relu & relu &  \\
Dense Units (Layer 2) & 64 & 64 & 32 & 64 & 32 &  &  & 64 & 128 & 32 &  \\
Activation (Dense 2) & sigmoid & relu & tanh & sigmoid & tanh &  &  & relu & relu & sigmoid &  \\
Dense Units (Layer 3) & 64 & 32 &  & 64 & 32 &  &  & 32 & 64 & 64 &  \\
Activation (Dense 3) & relu & relu &  & tanh & relu &  &  & sigmoid & tanh & tanh &  \\
\bottomrule
\end{tabular}
\end{adjustbox}
 \vspace{0.5em}
{\scriptsize
\noindent\parbox{\textwidth}{
Note: The table presents the selected hyperparameter configuration for the LSTM model in each rolling window, chosen after a randomized hyperparameter search. The tuning was performed using the Keras Tuner library with 25 executions and 3 repeated trials per execution, optimizing for the lowest average validation loss (MSE).
}
}
\end{table}

\subsubsection{LSTM Volatility Forecasting Methodology}

The LSTM volatility forecasting methodology leverages a structured approach to predict one-day-ahead volatility for the S\&P 500, utilizing log returns as inputs and producing volatility quantiles as outputs within a rolling window framework. Once optimal hyperparameters are selected (as detailed in Subsection 4.2.1), the LSTM model is trained independently within each of the 11 rolling windows, each comprising a training period of 11 years (2,772 trading days), a validation period of 3 years (756 days), and a test period of 1 year (252 days). During training, log returns are fed into the model, which processes them through its gated architecture to forecast the subsequent day’s volatility (\( t+1 \)), with continuous monitoring of validation loss—measured via mean squared error (MSE)—to ensure generalization to unseen data. An early stopping criterion halts training if the validation loss fails to improve over 10 consecutive epochs within a maximum of 100 epochs per window, retaining the parameters from the epoch with the lowest MSE. Following training, the model generates out-of-sample volatility predictions for the 252-day test period in each window, aggregating these forecasts across all windows to form a comprehensive set spanning January 1, 2000, to January 1, 2025. This complete forecast series enables robust evaluation against \chg{historical} volatility and benchmark models, assessing the LSTM’s effectiveness across diverse market conditions. \chg{The detailed workflow is presented in Algorithm 1.}

\begin{algorithm}
\caption{Volatility Forecasting Pipeline with LSTM and Hyperparameter Optimization}
\footnotesize  % Make the font smaller for the whole algorithm
\begin{algorithmic}[1]
\State \textbf{Initialize environment:} import required libraries and load S\&P 500 data.
\Statex \hspace{0.4cm} \texttt{import numpy, pandas, tensorflow, keras, keras\_tuner, sklearn}
\Statex \hspace{0.4cm} \texttt{df $\gets$ load\_SP500\_data()}
\State \textbf{Compute log returns and rolling volatility:}
\Statex \hspace{0.4cm} $\texttt{log\_return}_t \gets \log\left(\frac{\texttt{Close}_t}{\texttt{Close}_{t-1}}\right)$
\Statex \hspace{0.4cm} $\texttt{volatility}_t \gets \texttt{RollingStd}(\texttt{log\_return}, \texttt{window}=21)$
\State \textbf{Segment data using a rolling window approach:}
\For{each rolling window}
    \State \texttt{train, val, test $\gets$ split\_data(df, train=11y, val=3y, test=1y)}
    \State \texttt{X\_train, y\_train $\gets$ create\_sequences(train, lookback=21)}
    \State \texttt{X\_val, y\_val $\gets$ create\_sequences(val, lookback=21)}
    \State \texttt{X\_test, y\_test, test\_dates $\gets$ create\_sequences(test, lookback=21)}
    \State \texttt{scale(X\_train, X\_val, X\_test)}
\EndFor
\State \textbf{Define model and hyperparameter space}
\State \textbf{Run hyperparameter tuning:}
\State \texttt{tuner $\gets$ RandomSearch(build\_model, executions\_per\_trial=3, max\_trials=25)}
\State \texttt{tuner.search(X\_train, y\_train, ...)}
\State \textbf{Select best model with lowest average validation loss:}
\State \texttt{best\_model $\gets$ tuner.get\_best\_models(1)[0]}
\State \textbf{Train best model with early stopping:}
\State \texttt{early\_stop $\gets$ EarlyStopping(patience=5)}
\State \texttt{best\_model.fit(X\_train, y\_train, val\_data=(X\_val, y\_val), callbacks=[early\_stop])}
\State \textbf{Forecast on test data:}
\State \texttt{forecast $\gets$ best\_model.predict(X\_test)}
\State \texttt{predictions.append(forecast[0])}
\State \textbf{Aggregate results:}
\State \texttt{full\_forecast $\gets$ concat(predictions)}
\State \textbf{Evaluate forecast performance:}
\Statex \hspace{0.4cm} $\texttt{MSE} = \sqrt{\texttt{mean\_squared\_error}(y_{\text{true}}, \texttt{full\_forecast})}$
\Statex \hspace{0.4cm} $\texttt{MAE} = \texttt{mean\_absolute\_error}(y_{\text{true}}, \texttt{full\_forecast})$
\Statex \hspace{0.4cm} $\texttt{MAPE} = \texttt{mean\_absolute\_percentage\_error}(y_{\text{true}}, \texttt{full\_forecast})$
\State \textbf{Return:} \texttt{full\_forecast}, \texttt{MSE}, \texttt{MAE}, \texttt{MAPE}
\end{algorithmic}
\end{algorithm}

\chg{The model was implemented using Python 3.12.6 within Jupyter notebooks on Visual Studio Code. The software environment included TensorFlow and Keras for model architecture and training, with Keras Tuner employed for hyperparameter optimization using its random search functionality. 

The calculations were carried out on a standard consumer-grade laptop equipped with a multi-core CPU and 16 GB of RAM, without access to dedicated GPU acceleration. As a result, all computations were carried out on the CPU, which prolonged the processing time compared to GPU-supported environments.

Due to the repeated training required across rolling windows and the comprehensive nature of the hyperparameter tuning (performed separately on each training-validation window split), the tuning process spanned approximately 24 hours in total. Once optimal parameters were determined, the training of the LSTM model and subsequent forecasting per window were significantly faster, requiring roughly 15 minutes each. Despite hardware limitations, the sequential design and use of efficient libraries ensured the feasibility of model development and evaluation.}

\subsubsection{LSTM Model Results}

The Long Short-Term Memory (LSTM) model predicts S\&P 500 \chg{21-day rolling historical volatility} for day (\( t+1 \)) from January 24, 2014, to December 30, 2024, using log returns. 

\begin{figure}[h!]
    \centering
    \caption{LSTM Predictions vs. \chg{21-Day Rolling Historical Volatility}}
    \includegraphics[width=1\linewidth]{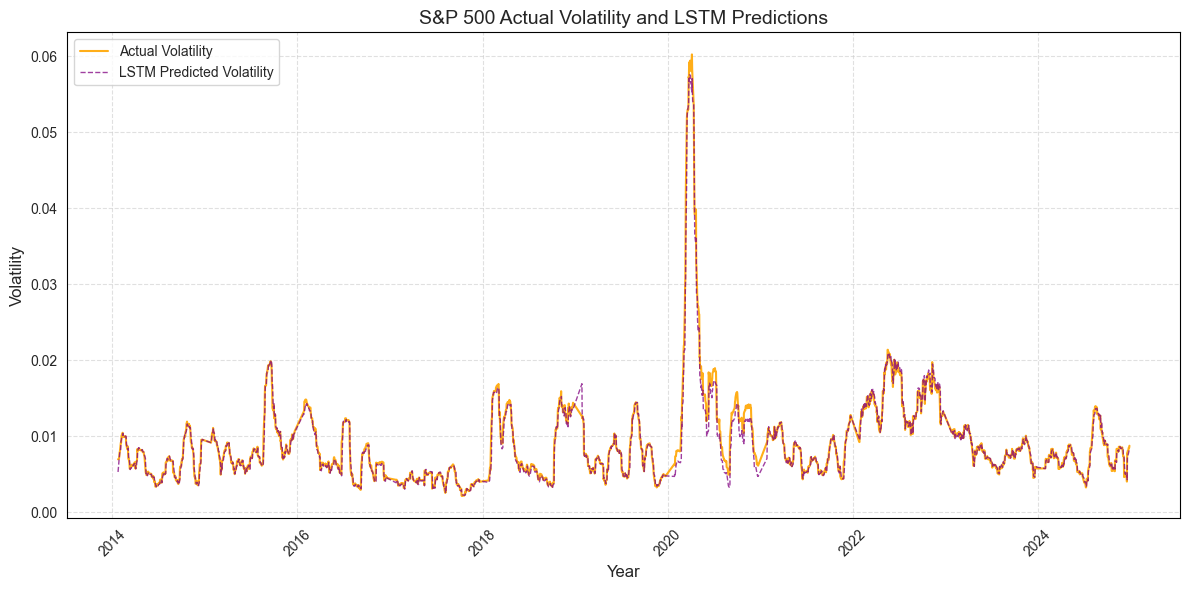}
    \label{fig:lstm}

    \vspace{0.5em}
{\scriptsize
\noindent\parbox{\textwidth}{
Note: The figure shows the predicted rolling volatility for day \(t+1\), provided by LSTM model, plotted against the \chg{21-day rolling historical} volatility for the period from January 2014 to December 2024.
}
}
\end{figure}

Figure~\ref{fig:lstm} compares these out-of-sample forecasts to the 21-day rolling volatility, showing effective trend capture but underestimation of sharp 2020–2022 shifts.

Table~\ref{tab:error_metrics2} lists error metrics: MAPE of 5.29\% indicates strong relative accuracy, with MSE of \( 7.09 \times 10^{-7} \) and MAE of \( 4.80 \times 10^{-4} \) showing small absolute errors.

\begin{table}[h!]
\centering
\caption{Comparative Error Metrics: LSTM Model vs. \chg{21-Day Rolling Historical Volatility} on Period January 2014 - December 2024}
\label{tab:error_metrics2}
\begin{tabular}{lc}
\hline
\textbf{Metric} & \textbf{Value} \\
\hline
Mean Absolute Percentage Error (MAPE) & 5.29\% \\
Mean Squared Error (MSE) & \( 7.09 \times 10^{-7}\)  \\
Mean Absolute Error (MAE) & \(4.80 \times 10^{-4}\) \\
\hline
\end{tabular}

\vspace{0.5em}
{\scriptsize
\noindent\parbox{\textwidth}{
Note: The table presents the accuracy metrics calculated for the out-of-sample predictions of the LSTM model for day \(t+1\), covering the period from January 2014 to December 2024.
}
}
\end{table}

\subsection{Hybrid SV-LSTM Model}

The hybrid SV-LSTM model integrates predictions from the Stochastic Volatility (SV) model as an additional input to the Long Short-Term Memory (LSTM) framework, combining the probabilistic strengths of SV with the LSTM’s capacity to capture complex temporal dependencies. This approach aims to improve the accuracy and robustness of S\&P 500 volatility forecasting beyond the standalone SV and LSTM models, described in more detail in Sections 4.1 and 4.2, respectively. 

By leveraging both stochastic and sequential modeling, the hybrid targets the 21-day rolling historical volatility for day (\( t+1 \)), evaluated over January 24, 2014, to December 30, 2024.

\subsubsection{SV-LSTM Model Framework}

The hybrid model utilizes three inputs: log returns, 21-day rolling \chg{historical} volatility and the SV model’s latent volatility output \( \sqrt{e^{h_{t+1}}} \), where \( h_{t+1} \) is the log-volatility from posterior draws, as mentioned in Section 4.1. 

\chg{The SV component, trained on log returns from December 1998 to December 2024,
with a 504-day rolling window, generated the prediction of latent estimates for day (t+1) and is refitted at each timestep. As the SV estimation is based on the past close prices for the last two years, the prediction for day (t+1) is obtained at the same timestep as we have the close price for the respective day t. This allowed to fetch the estimates with the input data for LSTM model for day t, such as log returns and rolling volatility estimates. }

Before feeding the data into the model, all inputs undergo preprocessing, including scaling and segmentation into 21-day sequences, following the procedure detailed in Section 3.3. The hybrid model employs the same LSTM architecture described in Section 4.2 and explores the same hyperparameter space defined in Table ~\ref{tab:lstm_hyperparams}. 

Hyperparameter tuning is performed independently within each rolling window using a random search strategy over 25 trials, with 3 repetitions per trial to account for training variability. The results are averaged based on the Mean Squared Error (MSE) to select the optimal configuration.

Model training and evaluation are conducted across 11 overlapping rolling windows covering the period from 2000 to 2024. For each window, the model is trained on the most recent 504 days and generates forecasts for the next day 
(\( t+1 \)) utilizing the preceding 21-day sequences.

\subsubsection{SV-LSTM Model Results}

\begin{figure}[h!]
    \centering
     \caption{SV-LSTM predictions plotted against actual volatility on out-of-sample values}
    \includegraphics[width=1\linewidth]{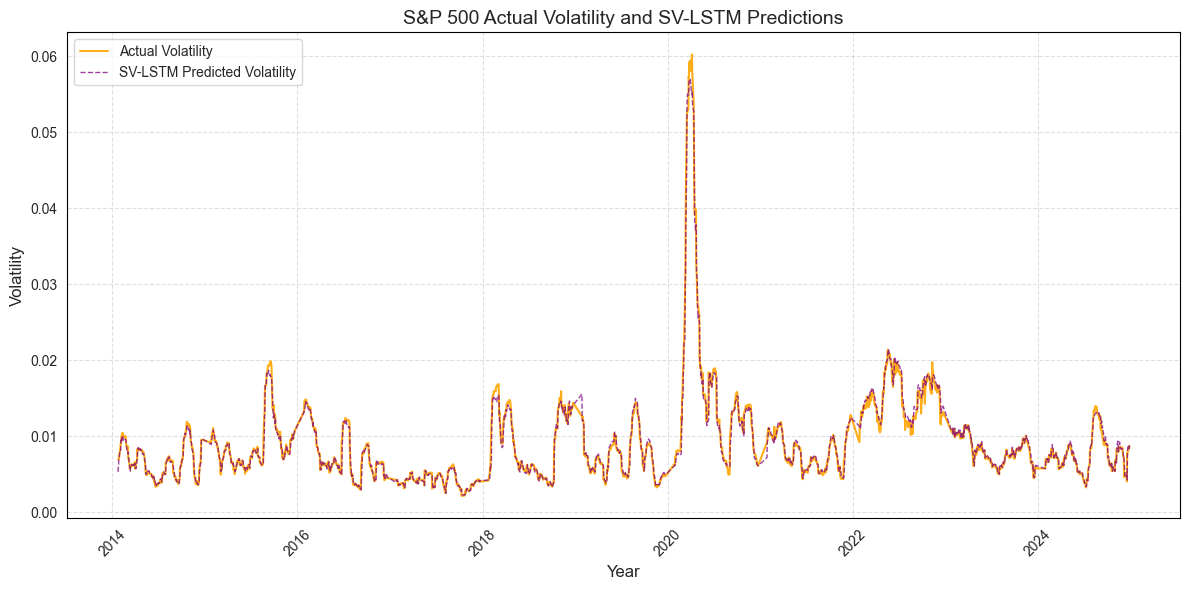}
    \label{fig:sv-lstm}

\vspace{0.5em}
{\scriptsize
\noindent\parbox{\textwidth}{
Note: The table presents the accuracy metrics calculated for the out-of-sample predictions of the SV-LSTM model for day \(t+1\), covering the period from January 2014 to December 2024.
}
}    
\end{figure}

The hybrid SV-LSTM model predicts S\&P 500 \chg{21-day rolling historical} volatility for day (\( t+1 \)) from January 24, 2014, to December 30, 2024, using out-of-sample data concatenated from the test segments of the 11 sliding windows. Figure~\ref{fig:sv-lstm} plots these predictions against the actual 21-day rolling volatility, demonstrating effective capture of overall trends and superior adaptability compared to the standalone LSTM (Section 4.2.3), particularly during the sharp 2020–2022 shifts driven by the COVID-19 pandemic.

Table~\ref{tab:error_metrics3} presents the comparative error metrics against the \chg{21-day rolling historical} volatility benchmark. The Mean Absolute Percentage Error (MAPE) of 4.75\% indicates high relative accuracy and reflects the hybrid’s ability to closely track proportional volatility changes. This is supported by a Mean Squared Error (MSE) of \( 5.07 \times 10^{-7} \) and Mean Absolute Error (MAE) of \( 4.29 \times 10^{-4} \).

\begin{table}[h!]
\centering
\caption{Comparative Error Metrics: SV-LSTM Model vs. \chg{21-Day Rolling Historical Volatility} on Period January 2014 - December 2024}
\label{tab:error_metrics3}
\begin{tabular}{lc}
\hline
\textbf{Metric} & \textbf{Value} \\
\hline
Mean Absolute Percentage Error (MAPE) & 4.75\% \\
Mean Squared Error (MSE) & \( 5.07 \times 10^{-7}\)  \\
Mean Absolute Error (MAE) & \(4.29 \times 10^{-4}\) \\
\hline
\end{tabular}

\vspace{0.5em}
{\scriptsize
\noindent\parbox{\textwidth}{
Note: The table presents the accuracy metrics calculated for the out-of-sample predictions of the SV-LSTM model for day \(t+1\), covering the period from January 2014 to December 2024.
}
}
\end{table}

\section{Empirical Results and Benchmark Comparison}

\subsection{Benchmark Comparison}

This section synthesizes and compares the performance of all 3 models described previously in subsections 4.1.3, 4.2.3, 4.3.2 of their results of predicting S\&P 500 rolling volatility on out-of-sample values on period from January 2014 until December 2024. The performance metrics along all three models are \chg{presented} together in the Table ~\ref{tab:metrics_all}. As we can observe, the hybrid SV-LSTM model surpasses the benchmark models along all three metrics, highlighting the leverage success.

\begin{table}[h!]
    \centering
    \caption{Error Metrics for SV, LSTM, and Hybrid SV-LSTM Models vs. \chg{21-Day Rolling Historical Volatility} (January 2014 – December 2024)}
    \label{tab:metrics_all}
    \begin{tabular}{lccc}
        \hline
        \textbf{Model} & \textbf{MAPE (\%)} & \textbf{MSE} & \textbf{MAE} \\
        \hline
        SV & 18.12 & \( 9 \times 10^{-6} \) & \( 1.717 \times 10^{-3} \) \\
        LSTM & 5.29 & \( 7.09 \times 10^{-7} \) & \( 4.80 \times 10^{-4} \) \\
        Hybrid SV-LSTM & 4.75 & \( 5.07 \times 10^{-7} \) & \( 4.29 \times 10^{-4} \) \\
        \hline
    \end{tabular}

    \vspace{0.5em}
{\scriptsize
\noindent\parbox{\textwidth}{
Note:  The table presents the accuracy metrics calculated for out-of-sample predictions generated by three models: SV, LSTM, and the hybrid SV-LSTM. These predictions target day \(t+1\) and cover the period from January 2014 to December 2024.
}
}
\end{table}

\subsection{Statistical Tests}

The hybrid SV-LSTM model’s performance was compared to LSTM and SV models using the Wilcoxon Signed-Rank Test and Diebold-Mariano (DM) Test, assessing forecast error differences (\( e_t = \hat{y}_t - y_t \)) \chg{between tested models} from January 24, 2014, to December 30, 2024.

The Wilcoxon test, \chg{firstly proposed by} \textcite{wilcoxon1945individual}, checks if two models’ error distributions are similar without assuming a specific shape, identifying its non-parametric feature. It tests the null hypothesis (\( H_0 \)): the median error is the same for both models, against the alternative (\( H_1 \)): the medians differ. We calculate differences between paired errors, rank them, and compute a statistic to see if the difference is significant.

\begin{equation}
d_t = e_{1,t} - e_{2,t}
\end{equation}
Where: 

- \( d_t \): Difference in errors at time \( t \) between two models.

- \( e_{1,t} \): Error of the first model at time \( t \), where \( e_{1,t} = \hat{y}_{1,t} - y_t \) (\( \hat{y}_{1,t} \) is the predicted value, \( y_t \) is the actual value).

- \( e_{2,t} \): Error of the second model at time \( t \).

These differences (\( d_t \)) are ranked by their absolute size (\( |d_t| \)), ignoring zeros, and given their original signs (positive or negative). The test statistic \( W \) is the sum of positive ranks. For many observations (large \( n \)), we standardize \( W \) into a \( Z \)-score:

\begin{equation}
Z = \frac{W - \frac{n(n+1)}{4}}{\sqrt{\frac{n(n+1)(2n+1)}{24}}}
\end{equation}

Where: 

- \( Z \): Standardized score, compared to a normal distribution to find the p-value.

- \( W \): Sum of ranks for positive \( d_t \) values.

- \( n \): Number of non-zero differences, ~2,500 days from 2014–2024).

- \( \frac{n(n+1)}{4} \): Expected value of \( W \) if \( H_0 \) is true (no difference).

- \( \sqrt{\frac{n(n+1)(2n+1)}{24}} \): Standard deviation of \( W \), measuring its variability.

If \( p > 0.05 \), we \chg{cannot reject} \( H_0 \) that states there is no difference in errors produced by two models; if \( p < 0.05 \), we reject it, which implies that \chg{compared} models differ in their error distribution.

The Diebold-Mariano(DM) test, \chg{proposed by} \textcite{diebold2002comparing}, checks which model forecasts more accurately over time. Its null hypothesis (\( H_0 \)) says both models are equally accurate, while the alternative (\( H_1 \)) suggests one is better. We measure accuracy with a loss function \( L \), like squared error (\( L(e) = e^2 \)) for MSE or absolute error (\( L(e) = |e| \)) for MAE, and compute differences:

\begin{equation}
d_t = L(e_{1,t}) - L(e_{2,t})
\end{equation}

- \( d_t \): Difference in loss ( squared or absolute errors) at time \( t \).

- \( L(e_{1,t}) \): Loss for the first model’s error at \( t \).

- \( L(e_{2,t}) \): Loss for the second model’s error at \( t \).

- \( e_{1,t}, e_{2,t} \): Errors as defined above.

Under \( H_0 \), the average difference over time should be zero:

\begin{equation}
E[d_t] = 0
\end{equation}
- \( E[d_t] \): Expected (average) loss difference across all \( t \). If true, models have equal accuracy.

For \( H_1 \), we test if Model 1 is more accurate than Model 2, expecting:

\begin{equation}
E[d_t] < 0
\end{equation}
- \( E[d_t] < 0 \): Average loss difference is negative, meaning errors of the Model 1 are smaller.

The DM statistic is calculated from \( d_t \), adjusted for time-series patterns, and gives a p-value. If \( p < 0.05 \), we reject \( H_0 \), concluding Model 1 is more accurate. Significance is checked at \( \alpha = 0.05 \).

Table~\ref{tab:stat_tests} summarizes the statistical test results for model comparisons over 2014–2024. 

\begin{table}[h!]
    \centering
    \caption{Statistical Test Results for Model Comparisons \chg{on Period January 2014 - December 2024}}
    \label{tab:stat_tests}
    \resizebox{\textwidth}{!}{%
    \begin{tabular}{lccccc}
        \hline
        \textbf{Comparison} & \multicolumn{2}{c}{\textbf{Wilcoxon}} & \multicolumn{3}{c}{\textbf{Diebold-Mariano}} \\
        \cline{2-6}
        & \textbf{\( W \) (\( Z \))} & \textbf{\( p \)-value} & \textbf{DM (MSE)} & \textbf{DM (MAE)} & \textbf{\( p \)-value} \\
        \hline
        LSTM vs. SV & 319,921 (2.75) & \( < 0.001 \) & -8.94 & -27.57 & \( < 0.001 \) \\
        SV-LSTM vs. LSTM & 1,544,740 (1.90) & 0.058 & -6.86 & -5.51 & \( < 0.001 \) \\
        SV-LSTM vs. SV & 266,472 (3.87) & \( < 0.001 \) & -9.18 & -29.14 & \( < 0.001 \) \\
        \hline
    \end{tabular}
    }
     \vspace{0.5em}
{\scriptsize
\noindent\parbox{\textwidth}{
Note: The table reports the results of the Wilcoxon signed-rank test and the Diebold-Mariano test used to compare the forecasting performance of the SV-LSTM model against the LSTM and SV models over the period from December 2014 to January 2024. The Wilcoxon test evaluates the statistical significance of differences in prediction accuracy, while the Diebold-Mariano test compares forecast errors based on MSE and MAE.
}
}
\end{table}

\chg{Comparing SV and LSTM, the Wilcoxon test statistic amounted to} \( W = 319{,}921 \) (\( Z = 2.75, \, p < 0.001 \)), allowing for the rejection of \( H_0 \) and indicating a significant difference in the error distributions between the two models. The Diebold-Mariano (DM) test statistics, \( DM = -8.94 \) (\( p < 0.001 \)) for MSE and \( DM = -27.57 \) (\( p < 0.001 \)) for MAE, also led to the rejection of \( H_0 \). These results identified the superior forecasting accuracy of the LSTM model compared to the SV model.

For the SV-LSTM vs. LSTM comparison, the Wilcoxon Signed-Rank Test obtained a statistic of \( W = 1,544,740 \) (\( Z = 1.90 \), \( p = 0.058 \)), driven by a large sample size (\( n \approx 2,500 \), 2014–2024). Although \( p = 0.058 > 0.05 \)\chg{ is close to the critical value}, we fail to reject \( H_0 \), indicating no statistically significant difference in the error distributions (\( e_t = \hat{y}_t - y_t \)) between the two models. 
Meanwhile, the Diebold-Mariano (DM) Test for MSE gives \( DM = -6.86 \) (\( p < 0.001 \)), and for MAE, \( DM = -5.51 \) (\( p < 0.001 \)), both rejecting \( H_0 \) and demonstrating SV-LSTM’s better forecasting accuracy, as negative DM values reflect lower squared and absolute errors for the hybrid model.

For SV-LSTM vs. SV, the Wilcoxon test obtained the test statistics such as \( W = 266,472 \) (\( Z = 3.87 \), \( p < 0.001 \)), rejecting \( H_0 \) and confirming a significant difference between medians of error distributions. The DM Test further supports this conclusion: for MSE, \( DM = -9.18 \) (\( p < 0.001 \)), and for MAE, \( DM = -29.14 \) (\( p < 0.001 \)), both rejecting \( H_0 \) and indicating SV-LSTM’s  better accuracy over the SV model, with large negative DM statistics underscoring substantial error reductions across both metrics.

\section{Sensitivity Analysis}

Sensitivity analysis is a crucial step in the analysis of the proposed model as it takes into account the other settings and propositions that could have been omitted in the base case scenario, as also provides more insights to the model performance, as well as the areas of potential improvements and implementations. 

\subsection{Sequence Length}

\begin{figure}[h!]
    \centering   
    \caption{SV-LSTM Predictions with the Sequence of 5 Days}
    \includegraphics[width=1\linewidth]{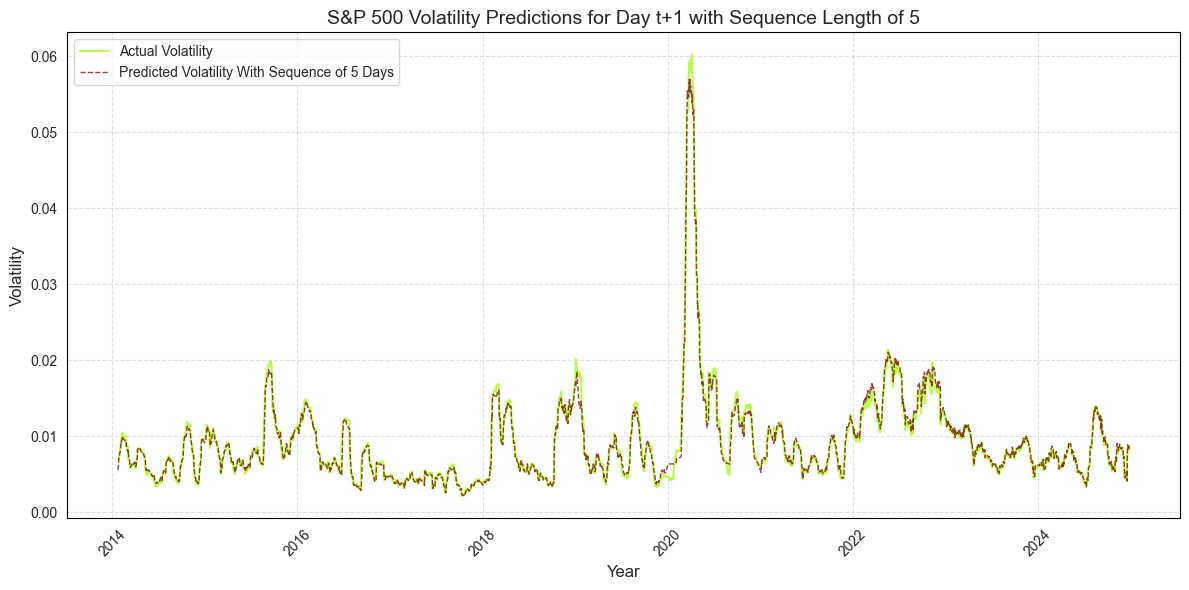}
    \label{fig:5seq}

    \vspace{0.5em}
{\scriptsize
\noindent\parbox{\textwidth}{
Note: The figure shows the predicted rolling volatility for day \(t+1\), provided by SV-LSTM model with a sequence of 5 days, plotted against the \chg{21-day rolling historical} volatility for the period from January 2014 to December 2024.
}
}
\end{figure}

A detailed analysis was conducted to investigate how varying the sequence length impacts the predictive performance of the model. In this context, a sequence length of five past observations was selected as input data for the hybrid SV-LSTM model, which is designed to forecast the \chg{21-day rolling historical} volatility for the following day (t+1). This specific configuration aims to capture short-term temporal dependencies in the data while maintaining model efficiency. The predicted values generated by the SV-LSTM model, plotted alongside the actual observed volatilities, are presented in Figure~\ref{fig:5seq}, providing a visual comparison of the model’s forecasting accuracy under this input setting.

Table~\ref{tab:5seq_metrics} presents a comparison of performance metrics between the modified hybrid model using a sequence length of 5 days and the baseline hybrid model with a sequence length of 21 days. The results indicate that reducing the sequence length does not lead to any improvement in the model’s predictive accuracy. In fact, the findings suggest that a longer input sequence may provide more stable and informative patterns for forecasting volatility.

\begin{table}[h!]
    \centering
    \caption{Error Metrics: Hybrid SV-LSTM with 5-Day Sequence vs. Baseline SV-LSTM (2014–2024)}
    \label{tab:5seq_metrics}
     \resizebox{\textwidth}{!}{
    \begin{tabular}{lcc}
        \hline
        \textbf{Metric} & \textbf{5-Day Sequence} & \textbf{Baseline SV-LSTM} \\
        \hline
        Mean Absolute Percentage Error (MAPE) & 5.47\% & 4.75\% \\
        Mean Squared Error (MSE) & \( 6.13 \times 10^{-7} \) & \( 5.07 \times 10^{-7} \) \\
        Mean Absolute Error (MAE) & \( 4.82 \times 10^{-4} \) & \( 4.29 \times 10^{-4} \) \\
        \hline
    \end{tabular}
    }
    \vspace{0.5em}
{\scriptsize
\noindent\parbox{\textwidth}{
Note: The table presents the accuracy metrics for out-of-sample predictions generated by the SV-LSTM model using a 5-day input sequence for forecasting day \(t+1\). These results are compared to those of the baseline hybrid model, which uses a 21-day input sequence. The evaluation covers the period from January 2014 to December 2024.
}
}
\end{table}

\begin{figure}[h!]
    \centering   
    \caption{SV-LSTM Predictions with the Sequence of 42 Days}
    \includegraphics[width=1\linewidth]{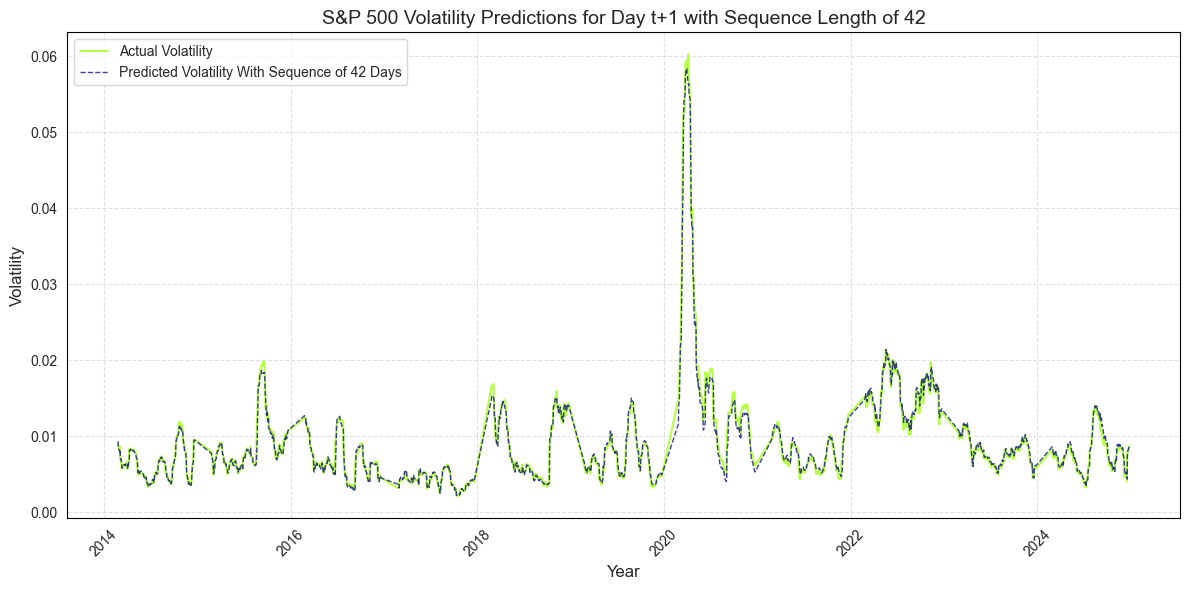}
    \label{fig:42seq}

    \vspace{0.5em}
{\scriptsize
\noindent\parbox{\textwidth}{
Note: The figure shows the predicted rolling volatility for day \(t+1\), provided by SV-LSTM model with a sequence of 42 days, plotted against the \chg{21-day rolling historical} volatility for the period from January 2014 to December 2024.
}
}
\end{figure}

\chg{Furthermore, another analysis was conducted to examine the influence of prolonging the input sequence into the SV-LSTM model to 42 days of past data. This analysis was aimed at examining the influence of providing more information and fluctuations, compared to the baseline model with 21 days of past data, suggesting that a longer period of observations could provide a more accurate forecast for day (t+1). In Figure}~\ref{fig:42seq} \chg{the predicted values generated by the SV-LSTM model are plotted alongside the actual observed volatilities, providing a visual comparison of the model’s forecasting accuracy under this input setting.}

Table~\ref{tab:5seq_metrics} \chg{presents a comparison of performance metrics between the modified hybrid model using a sequence length of 42 days and the baseline hybrid model with a sequence length of 21 days. The increased values of the error metrics for the 42-day sequence variation suggest that the increase in the number of days in the sequence does not improve the accuracy of the model forecast.}

\begin{table}[h!]
    \centering
    \caption{Error Metrics: Hybrid SV-LSTM with 42-Day Sequence vs. Baseline SV-LSTM (2014–2024)}
    \label{tab:42seq_metrics}
     \resizebox{\textwidth}{!}{
    \begin{tabular}{lcc}
        \hline
        \textbf{Metric} & \textbf{42-Day Sequence} & \textbf{Baseline SV-LSTM} \\
        \hline
        Mean Absolute Percentage Error (MAPE) & 5.31\% & 4.75\% \\
        Mean Squared Error (MSE) & \( 5.68 \times 10^{-7} \) & \( 5.07 \times 10^{-7} \) \\
        Mean Absolute Error (MAE) & \( 4.65 \times 10^{-4} \) & \( 4.29 \times 10^{-4} \) \\
        \hline
    \end{tabular}
     }
    \vspace{0.5em}
{\scriptsize
\noindent\parbox{\textwidth}{
Note: The table presents the accuracy metrics for out-of-sample predictions generated by the SV-LSTM model using a 42-day input sequence for forecasting day \(t+1\). These results are compared to those of the baseline hybrid model, which uses a 21-day input sequence. The evaluation covers the period from January 2014 to December 2024.
}
}
\end{table}

\chg{The results of the analysis indicate that the amount of past days used as an input for forecasting the volatility for day (t+1) for the SV-LSTM model influences the prediction accuracy. Nevertheless, the findings suggest that the sequence of 21 past observations, utlized in the baseline model is the optimal one as the error metrics increase for both options: for increase of sequence to 42 days and decrease of it to 5 days.}

\subsection{Data Scaling}

\begin{figure}[h!]
    \centering 
    \caption{SV-LSTM Model Predictions With Standard Scaler}
    \includegraphics[width=1\linewidth]{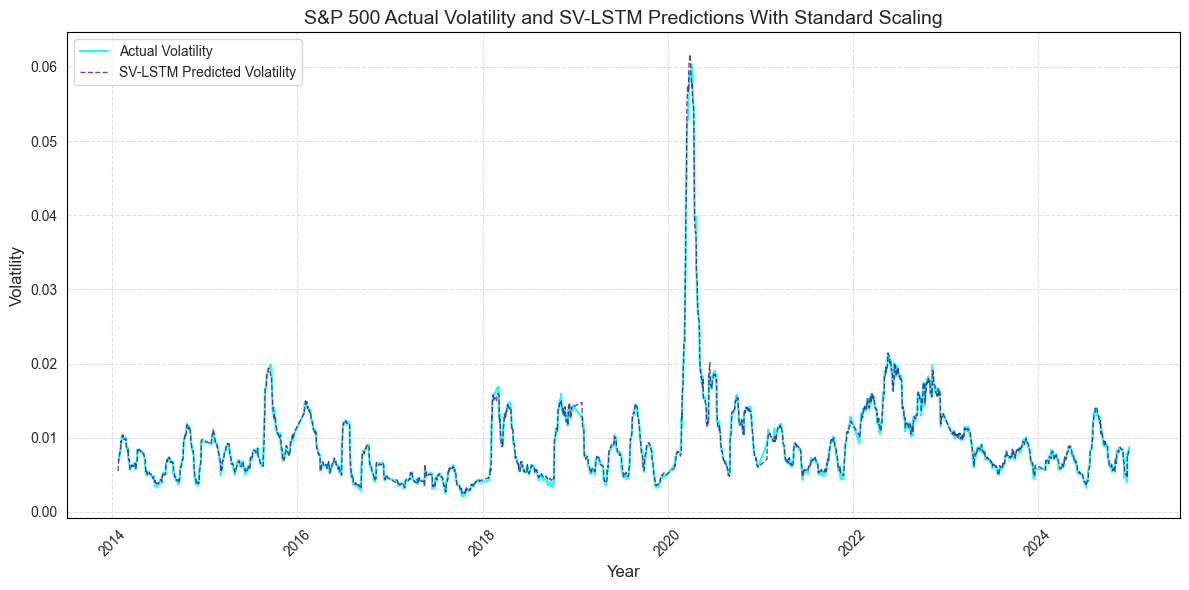}
    \label{fig:standrdsc}

     \vspace{0.5em}
{\scriptsize
\noindent\parbox{\textwidth}{
Note: The figure shows the predicted rolling volatility for day \(t+1\), provided by SV-LSTM model using a standard scaling preprocessing approach, plotted against the \chg{21-day rolling historical} volatility for the period from January 2014 to December 2024.
}
}
\end{figure}

Given the sensitivity of machine learning models to the quality and structure of input data, an additional analysis was conducted to examine the potential impact of data preprocessing on the performance of the hybrid model. Specifically, this analysis explored whether applying feature scaling could enhance prediction accuracy. Unlike the baseline hybrid model, where raw data was fed into the LSTM component using MinMax Scaler, in this modified version, the input data was standardized using a Standard Scaler, transforming the features to have zero mean and unit variance. The resulting predictions from the standardized-input hybrid model, plotted alongside the\chg{21-day rolling historical} volatility, are presented in Figure~\ref{fig:standrdsc}. 

Meanwhile, the corresponding accuracy metrics were computed and are presented in Table~\ref{tab:scaling_metrics}, alongside the results of the baseline SV-LSTM model without scaling. As noted, applying standard scaling to the input data led to a consistent improvement across all error metrics. Specifically, the Mean Absolute Percentage Error (MAPE) decreased from 4.75\% to 4.54\%, while the Mean Squared Error (MSE) and Mean Absolute Error (MAE) improved from \(5.07 \times 10^{-7}\) and \(4.29 \times 10^{-4}\) to \(4.18 \times 10^{-7}\) and \(3.80 \times 10^{-4}\), respectively. These results suggest that standardising the input data positively impacts the predictive accuracy of the hybrid SV-LSTM model.

\begin{table}[h!]
    \centering
    \caption{Error Metrics: SV-LSTM with Standard Scaling vs. Baseline SV-LSTM (2014–2024)}
    \label{tab:scaling_metrics}
     \resizebox{\textwidth}{!}{
    \begin{tabular}{lcc}
        \hline
        \textbf{Metric} & \textbf{SV-LSTM with Standard Scaling} & \textbf{Baseline SV-LSTM} \\
        \hline
        Mean Absolute Percentage Error (MAPE) & 4.54\% & 4.75\% \\
        Mean Squared Error (MSE) & \( 4.18 \times 10^{-7} \) & \( 5.07 \times 10^{-7} \) \\
        Mean Absolute Error (MAE) & \( 3.80 \times 10^{-4} \) & \( 4.29 \times 10^{-4} \) \\
        \hline
    \end{tabular}
     }
     \vspace{0.5em}
{\scriptsize
\noindent\parbox{\textwidth}{
Note: The table presents the accuracy metrics for out-of-sample predictions generated by the SV-LSTM model, which uses the StandardScaler approach for data preprocessing to forecast day  \(t+1\). These results are compared to those of the baseline hybrid model, which uses a MinMax scaling method.  The evaluation covers the period from January 2014 to December 2024.
}
}
\end{table}

Furthermore, another analysis was conducted where the raw input data into the LSTM model was preprocessed using the RobustScaler  transformation technique. The results were plotted against the actual values of rolling volatility across 2014-2024, as illustrated in Figure~\ref{tab:scaling_metrics_robust}. 

\begin{figure}[h!]
    \centering  
     \caption{SV-LSTM Model Predictions With Robust Scaler}
    \includegraphics[width=1\linewidth]{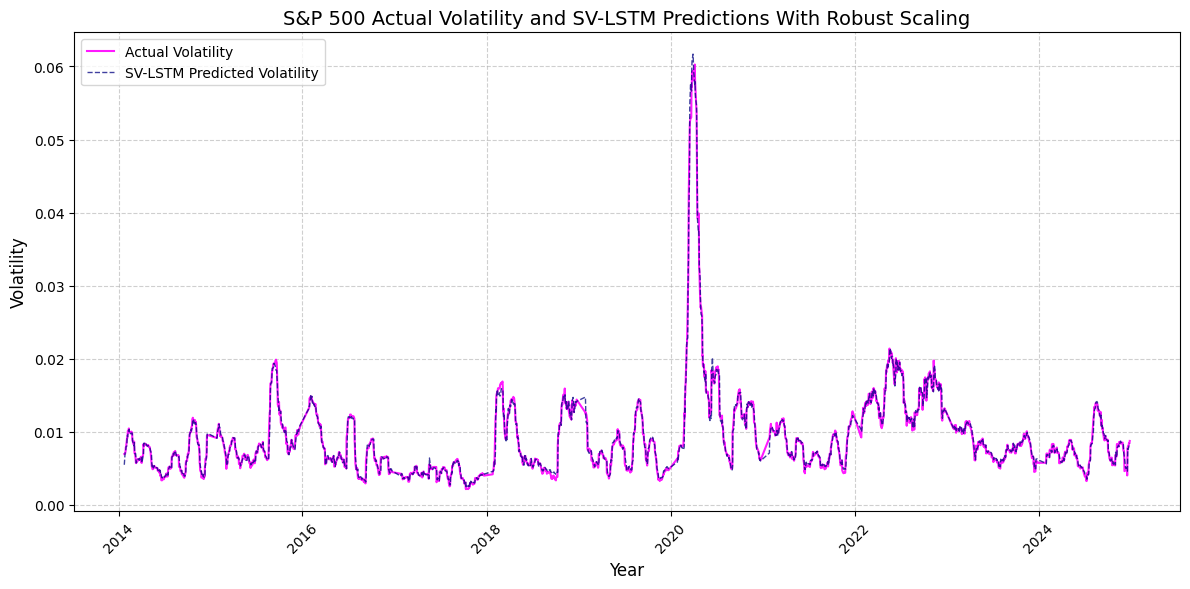}
    \label{fig: robust}

         \vspace{0.5em}
{\scriptsize
\noindent\parbox{\textwidth}{
Note: The figure shows the predicted rolling volatility for day \(t+1\), provided by SV-LSTM model using a robust scaling preprocessing approach, plotted against the \chg{21-day rolling historical} volatility for the period from January 2014 to December 2024.
}
}
\end{figure}
The corresponding metrics are summarised in Table~\ref{tab:scaling_metrics} compares the baseline SV-LSTM model without scaling and the alternative utilising RobustScaler.  
The results indicate that the change of the scaling approach has improved the results of the SV-LSTM model for all error metrics.  The Mean Absolute Percentage Error (MAPE) decreased from 4.75\% to 4.58\%, while the Mean Squared Error (MSE) and Mean Absolute Error (MAE) improved from \(5.07 \times 10^{-7}\) and \(4.29 \times 10^{-4}\) to \(4.69 \times 10^{-7}\) and \(3.90 \times 10^{-4}\), respectively. Therefore, we can conduct that applying robust scaling in data preprocessing step also improves the model perfomance, but not that much as standardscaling.
\begin{table}[h!]
    \centering
    \caption{Error Metrics: SV-LSTM with Robust Scaling vs. Baseline SV-LSTM (2014–2024)}
    \label{tab:scaling_metrics_robust}
     \resizebox{\textwidth}{!}{
    \begin{tabular}{lcc}
        \hline
        \textbf{Metric} & \textbf{SV-LSTM with Robust Scaling} & \textbf{Baseline SV-LSTM} \\
        \hline
        Mean Absolute Percentage Error (MAPE) & 4.58\%& 4.75\% \\
        Mean Squared Error (MSE) & \( 4.69 \times 10^{-7} \)& \( 5.07 \times 10^{-7} \) \\
        Mean Absolute Error (MAE) & \( 3.90 \times 10^{-4} \)& \( 4.29 \times 10^{-4} \) \\
        \hline
    \end{tabular}
     }
     \vspace{0.5em}
{\scriptsize
\noindent\parbox{\textwidth}{
Note: The table presents the accuracy metrics for out-of-sample predictions generated by the SV-LSTM model, which uses the RobustScaler approach for data preprocessing to forecast day  \(t+1\). These results are compared to those of the baseline hybrid model, which uses a MinMax scaling method.  The evaluation covers the period from January 2014 to December 2024.
}
}
\end{table}

The consistent improvement across all metrics confirms that the choice of data transformation can significantly influence the performance of deep learning-based volatility prediction models.

\subsection{Number of Hidden Layers}

\begin{figure}[h!]
    \centering  
     \caption{SV-LSTM Model Predictions With Varying Hidden Layers}
    \includegraphics[width=1\linewidth]{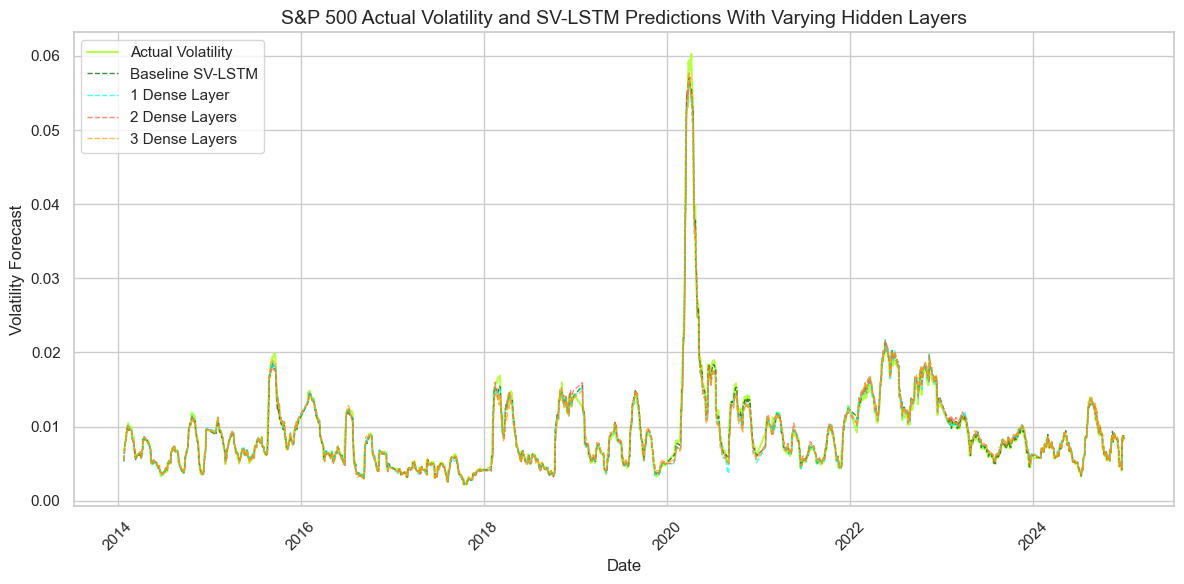}
    \label{fig:dense}

         \vspace{0.5em}
{\scriptsize
\noindent\parbox{\textwidth}{
Note: The figure presents the one-day-ahead rolling volatility forecasts generated by the SV-LSTM model with varying numbers of dense layers (1 to 3), compared against the 21-day rolling realized volatility of the S\&P 500 and the predictions of the baseline model, over the period from January 2014 to December 2024.
}
}
\end{figure}

The influence of differences in the structure of the LSTM model was also explored in the sensitivity analysis process. Specifically, the number of hidden layers was fixed in each scenario, ranging from 1 to 3 layers, and corresponding tuning, training, and forecasting were performed. Figure~\ref{fig:dense} represents all of the scenarios plotted against the baseline model and the actual volatility.

The corresponding metrics are summarized in Table~\ref{tab:metrics_dense}, which compares the baseline SV-LSTM model with the models having between 1 and 3 hidden layers. The results indicate that fixing the number of hidden layers downgrades the performance of the model, implying that flexible and dynamic tuning of the number of hidden layers in each rolling window produces better results.

\begin{table}[h!]
    \centering
    \caption{Error Metrics: SV-LSTM with Varying Dense Layers vs. Baseline SV-LSTM (2014–2024)}
    \label{tab:metrics_dense}
    \resizebox{\textwidth}{!}{
    \begin{tabular}{lcccc}
        \hline
        \textbf{Metric} & \textbf{1 Dense} & \textbf{2 Dense} & \textbf{3 Dense} & \textbf{Baseline SV-LSTM} \\
        \hline
        Mean Absolute Percentage Error (MAPE) & 5.73\% & 5.84\% & 6.40\% & 4.75\% \\
        Mean Squared Error (MSE) & \( 6.617 \times 10^{-7} \) & \( 6.961 \times 10^{-7} \) & \( 8.112 \times 10^{-7} \) & \( 5.07 \times 10^{-7} \) \\
        Mean Absolute Error (MAE) & \( 5.05 \times 10^{-4} \) & \( 5.24 \times 10^{-4} \) & \( 5.79 \times 10^{-4} \) & \( 4.29 \times 10^{-4} \) \\
        \hline
    \end{tabular}
    }
    \vspace{0.5em}
    {\scriptsize
    \noindent\parbox{\textwidth}{
    Note: The table presents the error metrics for the SV-LSTM model with varying numbers of dense layers (1–3) and compares them to the baseline SV-LSTM model. The evaluation covers the period from January 2014 to December 2024.
    }
    }
\end{table}

\section{Investment Strategy Simulation}

The selected data consisted of VIX monthly futures contracts, collected from the CBOE website for the entire testing period, from January 2014 to December 2024. The dataset included the trade date of each selected monthly future along with the open, high, low, close, settlement price, daily change, and total volume. For our strategy dataframe, selected values were the trade date, the respective monthly future symbol, and the close price.

Furthermore, the settlement dates and settlement values were downloaded for the entire testing period, again from January 2014 to December 2024. Each of these yearly datasets included the VIX monthly symbol, expiration date of each contract, and the close price on the expiration date.

To prepare the data for analysis, we combined the collected VIX futures data with the settlement dates and prices. We created a 'date to expiry' column, calculated as the difference between the expiration date and trade date. We then merged the forecasted volatility values (both actual smoothed rolling volatility and predicted values from day t-1) with the dataset. For the investment strategy, we moved the forecasted values by one day to align with the forecast for day t+1, ensuring that each entry in the dataframe corresponds to the correct day’s forecast.

The investment strategy is based on a daily comparison between the \textbf{forecasted volatility for day \( t+1 \)} and the \textbf{actual volatility observed at day \( t \)}. The trading rules are defined as follows:

\begin{itemize}
    \item \textbf{Long Position:} If the forecasted volatility \( \text{vol}_{t+1}^{\text{forecast}} \) is greater than the actual volatility \( \text{vol}_t^{\text{actual}} \), a \textbf{long signal} is generated (\( \text{signal} = 1 \)). This triggers a \textbf{long position in VIX futures}, entered at the close price of day \( t \).
    
    \item \textbf{Short Position:} If the forecasted volatility \( \text{vol}_{t+1}^{\text{forecast}} \) is less than the actual volatility \( \text{vol}_t^{\text{actual}} \), a \textbf{short signal} is generated (\( \text{signal} = -1 \)). This leads to a \textbf{short position in VIX futures}, also using the close price of day \( t \).
    
    \item \textbf{Position Management:} If the signal remains unchanged from the previous day (\( \text{signal}_t = \text{signal}_{t-1} \)), the position is maintained. If the signal changes (\( \text{signal}_t \ne \text{signal}_{t-1} \)), the current position is \textbf{closed}, and a \textbf{new position} is opened based on the updated signal. Transaction costs of \textbf{0.1\%} are applied at both entry and exit.
\end{itemize}

To evaluate the effectiveness of the trading strategy, it is compared against two naive benchmarks:

\begin{itemize}
    \item \textbf{ Long Only VIX Futures:} A strategy that maintains a constant long position throughout the entire period.
    \item \textbf{ Short Only VIX Futures:} A strategy that maintains a constant short position throughout the entire period.
\end{itemize}

The benchmarks ignore any information from volatility forecasts and serve to assess whether the model adds value beyond simple directional exposure.

\chg{For all strategies, a consistent assumption is utilised regarding the switching between futures contracts: each contract is held until its expiration, at which point the position is rolled into the next available monthly VIX future with the closest subsequent expiration date. The expiring position is closed at its official settlement price, therefore reflecting the return from holding the contract up until maturity. The new position is entered on the same day, using the close price of the next contract. This approach captures the final payoff of the expiring future while ensuring a smooth and consistent rollover mechanism across all strategy implementations.}

The strategy begins with an initial capital of \(\$1{,}000\). On each trading day, only 25\% of the available capital is allocated to the active position. Consequently, daily returns are scaled by a factor of 0.25 to reflect this partial investment policy.

\begin{figure}[h!]
    \centering  
     \caption{Equity Lines of Investment Strategy}
    \includegraphics[width=1\linewidth]{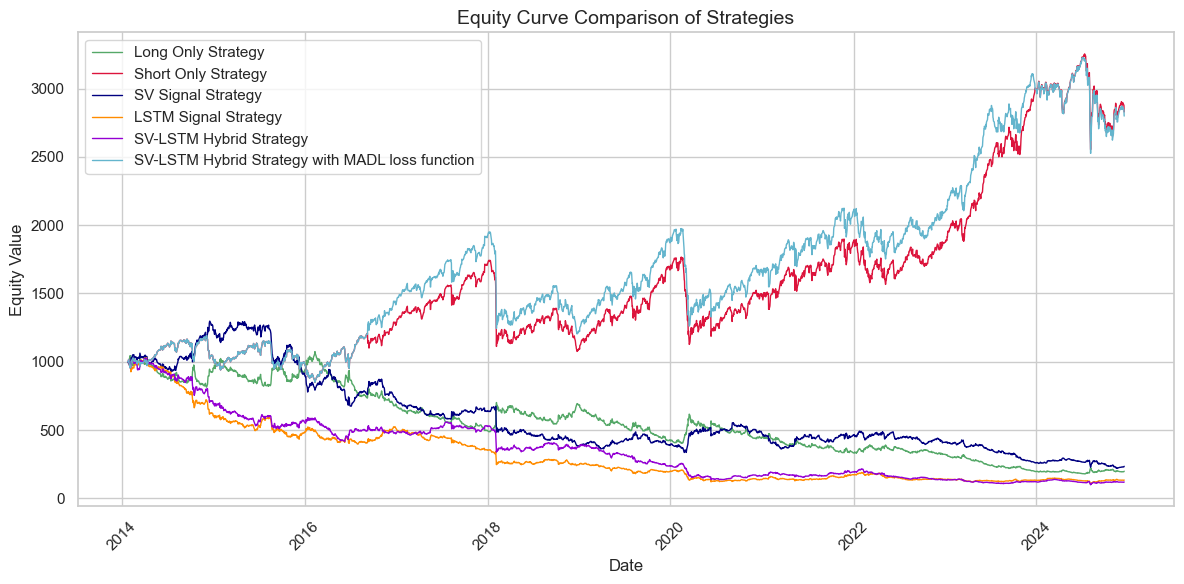}
    \label{fig:equity}

         \vspace{0.5em}
{\scriptsize
\noindent\parbox{\textwidth}{
Note: The figure presents the cumulative equity curves of five investment strategies based on VIX futures trading. The initial capital for each strategy is \$1,000, with 25\% allocated to each position daily. Transaction costs of 0.1\% are applied on both entry and exit.
}
}
\end{figure}

The equity lines of the strategies are shown in Figure~\ref{fig:equity}. The Short Only strategy outperformed all others, reflecting the inherent nature of VIX futures. During the volatile period at the start of 2020, the equity lines of both the LSTM and Short strategies declined significantly, while the SV strategy, driven by stochastic signals, gained substantially. It can be noted that the SV-LSTM strategy closely follows the SV line but with smoother fluctuations and higher overall equity than the LSTM.

The performance metrics presented in Table~\ref{tab:str_metrics} for the SV, LSTM, and SV-LSTM signal strategies indicate a pronounced underperformance. Except for Short Only strategy, both the highest Sharpe ratio of $-0.46$ and the annualised return of $-12.49\%$ were recorded for the SV Signal Strategy. These results highlight the necessity for further investigation into the neural network architectures to enhance the suitability of model forecasts for signal-based trading applications.

Given that the previous part of the study employed conventional error functions such as MSE, MAE, and MAPE, a modification to the loss function used during the compilation and training of the hybrid SV-LSTM model was introduced. The aforementioned error metrics measure discrepancies between forecasted and observed values, emphasising statistical accuracy rather than the predictive efficacy of the generated investment signals. Consequently, many studies optimise their models for statistical precision at the expense of economic profitability, thus neglecting the potential trading performance of the resulting signals.

To mitigate this limitation, the Mean Absolute Directional Loss (MADL), introduced by  ~\textcite{michankow2023madl}, was implemented. MADL is computed according to the following formula:
\begin{equation}
    \mathrm{MADL} = \frac{1}{N} \sum_{i=1}^{N} (-1) \times \mathrm{sign}(R_i \times \hat{R}_i) \times \mathrm{abs}(R_i),
\end{equation}
where $\mathrm{MADL}$ denotes the Mean Absolute Directional Loss, $R_i$ represents the observed return over interval $i$, $\hat{R}_i$ the predicted return in the same interval, $\mathrm{sign}(R_i \times \hat{R}_i)$ is the sign function returning values $(-1, 0, 1)$ based on the product $R_i \times \hat{R}_i$, $\mathrm{abs}(R_i)$ the absolute value of $R_i$, and $N$ the total number of forecasts. This approach explicitly links prediction direction to financial outcomes, enabling the model to assess whether a forecast would yield profit or loss, and to what degree.

\begin{table}[h!]
\centering
\caption{Performance Metrics of Equity Strategies}
\label{tab:str_metrics}
\resizebox{\textwidth}{!}{
\begin{tabular}{lcccccc}
\toprule
\textbf{Metric} & \textbf{Short Only} & \textbf{Long Only} & \textbf{SV} & \textbf{SV-LSTM} & \textbf{LSTM} & \textbf{SV-LSTM MADL} \\
\midrule
Sharpe Ratio      & 0.54 & -0.54 & -0.46 & -0.56 & -0.76 & 0.53\\
Calmar Ratio      & 0.26 & -0.17 & -0.15 & -0.17 & -0.20 & 0.26\\
Annualized Return (ARC)  & 10.03\% & -13.82\% & -12.49\% & -14.57\% & -18.31\% & 9.92\% \\
Annualized Std Dev (ASD) & 22.96\% & 22.96\% & 22.89\% & 23.07\% & 23.07\% & 22.95\% \\
Max Drawdown      & -38.27\% & -83.35\% & -83.11\% & -85.15\% & -93.37\% & -38.27\% \\
Total Return      & 183.02\% & -80.18\% & -76.60\% & -82.00\% & -88.94\% & 179.96\% \\
\bottomrule
\end{tabular}
}
         \vspace{0.5em}
{\scriptsize
\noindent\parbox{\textwidth}{
Note: This table presents key performance metrics for all evaluated strategies over the testing period from January 2014 to December 2024. The Sharpe Ratio and Annual Return (ARC) indicate risk-adjusted and average performance, respectively, while Annual Standard Deviation (ASD) reflects volatility, Max Drawdown captures the largest peak-to-trough equity decline, and Calmar Ratio measures risk-adjusted return relative to maximum drawdown.
}
}
\end{table}

All model parameters were held constant, with MADL introduced as the sole modification to the training configuration. The updated results revealed, shown in Table \ref{tab:str_metrics} that the SV-LSTM hybrid model, while still marginally underperforming relative to the baseline short-only strategy, exhibited improved coherence between predictive accuracy and investment performance. This finding suggests that incorporating the MADL loss function positively influences the model’s ability to capture directionally meaningful trading signals.

\section{Conclusion}

In this study, the hybrid SV-LSTM model was proposed and evaluated alongside benchmark models, namely its standalone components: the LSTM and the SV model.

\chg{The methodology of the introduced SV-LSTM model consisted of two parts. Firstly, predictions for day t+1 were generated by utilizing the statistical SV model, which was trained on log returns from December 1998 to January 2024 with a 504-day rolling window. This stage utilized daily close price for the S\&P 500 index, covering the period from January 1, 1998, to December 31, 2024, obtained from Yahoo Finance.

The second stage involved using the latent volatility estimates generated by SV model as inputs for the LSTM network. Since the SV estimation is based on the closing prices for the last two years, the prediction for day t+1 is obtained at the same timestep as the closing price for the respective day t. This allowed the latent estimates to be retrieved and treated as input data for the LSTM model for day t, along with logarithmic returns and 21-day rolling volatility estimates. In addition, the 21-day rolling volatility estimates were also set as the target for the LSTM model part. After the process of tuning and training, the LSTM part of the model produced the prediction for the 21-day rolling volatility of the S\&P 500 for day t+1.}

The error metrics demonstrated the superiority of the SV-LSTM model over both the LSTM and SV models. Additionally, the Diebold-Mariano (DM) test confirmed that the SV-LSTM model exhibits better predictive accuracy than both benchmark models. Furthermore, the Wilcoxon statistical tests indicated that the distribution of errors between the SV and SV-LSTM models is significantly different, whereas no significant difference was found between the SV-LSTM and LSTM models. The lack of the difference between the error distributions produced by both LSTM and SV-LSTM models can be explained by the shared LSTM core of prediction forecast as the last part of the model workflow, demonstrating shared limitations of the LSTM model that remained unresolved, identifying the ground for potential future research.

The investment strategy simulation further validated the practical relevance of the SV-LSTM model by assessing its performance on VIX futures data spanning from 2014 to 2024 in a realistic market environment. Given the underperformance of signal strategies based on models calibrated with conventional error metrics, the incorporation of the MADL loss function was undertaken. The results indicate that modifying the compilation loss function enhanced the signal strategy’s performance relative to the original Hybrid SV-LSTM model. This outcome underscores the importance of further research into loss functions that are better optimized for LSTM and other neural network models within applied financial contexts. Collectively, these findings highlight the value of integrating stochastic signals into neural architectures and suggest that future improvements could be achieved by refining neural network designs that balance classical statistical accuracy with practical trading effectiveness.

Exploring the sensitivity of the model to additional manipulations, it was found that shorter sequences reduce accuracy in opposition to the baseline of 21 days. This suggested that 21 days offers enough information for better predictability of the model, while the sequence of 5 days limits the amount of information and prevents the model from effectively learn short-term dynamics. Moreover, the data preprocessing part was also analyzed, and it was found that both standard and robust scaling improve the model forecast in contrast to the baseline min-max scaling, suggesting that data preprocessing options for machine learning models could be further explored.

The main limitation encountered in this work was the high computational cost of training each model, as every machine learning model underwent hyperparameter tuning across all rolling windows of training and validation data.

The following conclusions are drawn from the hypothesis testing conducted in this study:
\begin{itemize}
  \item H1: The results show that the inclusion of stochastic volatility forecasts for day t+1 enhances the predictive accuracy of the LSTM model.
  \item H2: Augmenting the input data of the LSTM model with external information beyond historical returns was proven to improve its forecasting performance.
  \item H3: The comparative metrics analysis showed that the hybrid SV-LSTM model delivers enhanced volatility forecasts compared to the standalone SV model.
\end{itemize}
In addition to the main hypotheses, the following secondary research questions underwent investigation:
\begin{itemize}
  \item RQ1  Increasing the dimensionality of inputs from the SV model further enhances the predictive performance of the hybrid model.
  \item RQ2 The change of data preprocessing scaling from min-max to either standard or robust scaling improved the performance of the SV-LSTM model.
  \item RQ3 The decreased sequence of the input data into the LSTM model was shown to not leverage the SV-LSTM prediction accuracy.

\end{itemize}

\chg{This study made a contribution to financial modelling by introducing and evaluating the combination of a statistical SV model as an input into a machine learning LSTM model. This approach allows the model to capture unpredictable market dynamics by incorporating the latent stochastic process and nonlinear dependencies in financial time series. The simulation of investment strategy alongside the sensitivity analysis provided practical insights into the application of the presented model.}

\chg{The hybrid SV-ML framework could be expanded in a number of ways in future studies. One approach involves feeding machine learning models with different outputs from the stochastic volatility model, such as latent states or predictive densities. Directly incorporating machine learning into the SV estimation procedure is another promising approach that could lessen the need for large-scale sampling by substituting effective approximators for computationally demanding simulations. Adaptability may also be improved by further research into more complex hybrid structures, such as fusing LSTM with more specialized SV models. Model stability may be increased by standardizing ML hyperparameters across windows.}

\printbibliography

@article{bahadori2019,
  author    = {Bahadori, M. T. and Lipton, Z. C.},
  title     = {Deep Learning for Financial Time Series Forecasting},
  journal   = {arXiv preprint arXiv:1807.02787}, 
  year      = {2019}
}

@article{blackscholes1973,
  author    = {Black, F. and Scholes, M.},
  title     = {The Pricing of Options and Corporate Liabilities},
  journal   = {Journal of Political Economy},
  year      = {1973},
  volume    = {81},
  number    = {3},
  pages     = {637--654},
  publisher = {University of Chicago Press},
  doi       = {10.1086/260062}
}

@article{bollerslev1986,
  author    = {Bollerslev, T.},
  title     = {Generalized Autoregressive Conditional Heteroskedasticity},
  journal   = {Journal of Econometrics},
  year      = {1986},
  volume    = {31},
  number    = {3},
  pages     = {307--327},
  publisher = {Elsevier},
  doi       = {10.1016/0304-4076(86)90063-1}
}

@article{breidt1998,
  author    = {Breidt, F. Jay and Crato, Nuno and de Lima, Pedro},
  title     = {The Detection and Estimation of Long Memory in Stochastic Volatility},
  journal   = {Journal of Econometrics},
  year      = {1998},
  volume    = {83},
  number    = {1-2},
  pages     = {325--348},
  publisher = {Elsevier},
  doi       = {10.1016/S0304-4076(97)00072-1}
}

@article{bucci2020,
  author    = {Bucci, A.},
  title     = {Realized Volatility Forecasting with Neural Networks},
  journal   = {Journal of Financial Econometrics},
  year      = {2020},
  volume    = {18},
  number    = {3},
  pages     = {502--531},
  publisher = {Oxford University Press},
  doi       = {10.1093/jjfinec/nbz009}
}

@article{diebold2002comparing,
  author    = {Diebold, F. X. and Mariano, R. S.},
  title     = {Comparing Predictive Accuracy},
  journal   = {Journal of Business \& Economic Statistics},
  year      = {2002},
  volume    = {20},
  number    = {1},
  pages     = {134--144},
  month     = {jan},
  publisher = {Taylor \& Francis, Ltd. on behalf of the American Statistical Association},
  doi       = {10.1198/073500102753410444}
}

@article{engle1982,
  author    = {Engle, R. F.},
  title     = {Autoregressive Conditional Heteroskedasticity with Estimates of the Variance of United Kingdom Inflation},
  journal   = {Econometrica},
  year      = {1982},
  volume    = {50},
  number    = {4},
  pages     = {987--1007},
  publisher = {Econometric Society},
  doi       = {10.2307/1912773}
}

@article{engle2001,
  author    = {Engle, R. F.},
  title     = {GARCH 101: The Use of ARCH/GARCH Models in Applied Econometrics},
  journal   = {Journal of Economic Perspectives},
  year      = {2001},
  volume    = {15},
  number    = {4},
  pages     = {157--168},
  publisher = {American Economic Association},
  doi       = {10.1257/jep.15.4.157}
}

@article{grudniewicz2023,
  author    = {Grudniewicz, S. and Slepaczuk, R.},
  title     = {Application of machine learning in algorithmic investment strategies on global stock markets},
  journal   = {Research in International Business and Finance},
  year      = {2023},
  volume    = {66},
  number    = {102052},
  publisher = {Elsevier},
  doi       = {10.1016/j.ribaf.2023.102052}
}

@article{heston1993,
  author    = {Heston, S. L.},
  title     = {A Closed-Form Solution for Options with Stochastic Volatility with Applications to Bond and Currency Options},
  journal   = {Review of Financial Studies},
  year      = {1993},
  volume    = {6},
  number    = {2},
  pages     = {327--343},
  publisher = {Oxford University Press},
  doi       = {10.1093/rfs/6.2.327}
}

@article{hochreiter1997,
  author    = {Hochreiter, S. and Schmidhuber, J.},
  title     = {Long Short-Term Memory},
  journal   = {Neural Computation},
  year      = {1997},
  volume    = {9},
  number    = {8},
  pages     = {1735--1780},
  publisher = {MIT Press},
  doi       = {10.1162/neco.1997.9.8.1735}
}

@article{kim1998,
  author    = {Kim, S. and Shephard, N. and Chib, S.},
  title     = {Stochastic Volatility: Likelihood Inference and Comparison with ARCH Models},
  journal   = {Review of Economic Studies},
  year      = {1998},
  volume    = {65},
  number    = {3},
  pages     = {361--393},
  publisher = {Oxford University Press},
  doi       = {10.1111/1467-937X.00050}
}

@article{kim2018,
  author    = {Kim, H. Y. and Won, C. H.},
  title     = {Forecasting the volatility of stock price index: A hybrid model integrating LSTM with multiple GARCH-type models},
  journal   = {Expert Systems with Applications},
  year      = {2018},
  volume    = {103},
  pages     = {25--37},
  publisher = {Elsevier},
  doi       = {10.1016/j.eswa.2018.03.002}
}

@article{kilic2011,
  author    = {Kilic, R.},
  title     = {Long memory and nonlinearity in conditional variances: A smooth transition FIGARCH model},
  journal   = {Journal of Empirical Finance},
  year      = {2011},
  volume    = {18},
  number    = {3},
  pages     = {368--378},
  publisher = {Elsevier},
  doi       = {10.1016/j.jempfin.2011.02.001}
}

@article{michankow2022,
  author    = {Michańków, J. and Sakowski, P. and Ślepaczuk, R.},
  title     = {LSTM in Algorithmic Investment Strategies on BTC and S\&P500 Index},
  journal   = {Sensors},
  year      = {2022},
  volume    = {22},
  number    = {3},
  pages     = {917},
  publisher = {MDPI},
  doi       = {10.3390/s22030917}
}

@article{nelson2001,
  author    = {Nelson, D. B.},
  title     = {Conditional Heteroskedasticity in Asset Returns: A New Approach},
  journal   = {Econometrica},
  year      = {1991},
  volume    = {59},
  number    = {2},
  pages     = {347--370},
  publisher = {Econometric Society},
  doi       = {10.2307/2938260}
}

@article{nguyen2019,
  title={A long short-term memory stochastic volatility model},
  author={Nguyen, N. and Tran, M. and Gunawan, D. and Kohn, R.},
  journal={arXiv preprint arXiv:1906.02884},
  pages={120},
  year={2019}
}

@article{rozynska2024,
  author    = {Roszyk, K. and {\'S}lepaczuk, R.},
  title     = {The Hybrid Forecast of S\&P 500 Volatility ensembled from VIX, GARCH and LSTM models},
  journal   = {arXiv preprint arXiv:2407.16780}, 
  year      = {2024},
}

@article{taylor1982,
  author    = {Taylor, S. J.},
  title     = {Financial Returns Modelled by the Product of Two Stochastic Processes - A Study of Daily Sugar Prices 1961-79},
  journal   = {Time Series Analysis: Theory and Practice},
  year      = {1982},
  volume    = {1},
  pages     = {203--226},
  publisher = {North-Holland},
}

@article{wilcoxon1945individual,
  author    = {Wilcoxon, F.},
  title     = {Individual Comparisons by Ranking Methods},
  journal   = {Biometrics Bulletin},
  year      = {1945},
  volume    = {1},
  number    = {6},
  pages     = {80--83},
  month     = {dec},
  publisher = {International Biometric Society},
  doi       = {10.2307/3001968}
}

@book{taylor1986,
  author    = {Taylor, S. J.},
  title     = {Modelling Financial Time Series},
  publisher = {John Wiley \& Sons},
  year      = {1986}
}

@article{yu2002,
  author    = {Yu, J.},
  title     = {Forecasting volatility in the New Zealand stock market},
  journal   = {Applied Financial Economics},
  year      = {2002},
  volume    = {12},
  number    = {3},
  pages     = {193--202},
  publisher = {Taylor & Francis},
  doi       = {10.1080/09603100110088094}
}

@article{michankow2023madl,
  author       = {Jakub Michańków and Paweł Sakowski and Robert Ślepaczuk},
  title        = {Mean Absolute Directional Loss as a New Loss Function for Machine Learning Problems in Algorithmic Investment Strategies},
  journal      = {arXiv preprint arXiv:2309.10546},
  year         = {2023},
  volume       = {abs/2309.10546},
  url          = {https://arxiv.org/abs/2309.10546},
  publisher    = {arXiv},
}

\end{document}